\documentclass[aps,pre,preprint,floatfix]{revtex4}

\usepackage{latexsym}
\usepackage[usenames]{color}
\usepackage{amsthm}
\usepackage{hyperref}

\usepackage{amsmath}    
\usepackage{amsfonts}  
\usepackage{amssymb}
\usepackage{graphicx}
\usepackage{slashed}
\usepackage{fouridx}

\theoremstyle{plain} 
\theoremstyle{plain} 
\theoremstyle{plain}  
\theoremstyle{plain} 
\theoremstyle{plain}  
\theoremstyle{remark} 
\theoremstyle{plain} 
\theoremstyle{remark}

\newcommand\CROSS[1]{%
  \hbox{%
    \vbox{
      \hrule
      \kern2.5pt
      \hbox{$#1$\,\strut}
    }%
  \vrule
  }\mskip\thickmuskip
}

\begin{document}
\begin{center}
\Large{\textbf{Semipredictable dynamical systems}}\\ 
~\\

\large{Vladimir Garc\'{\i}a-Morales}\\

\normalsize{}
~\\

Departament de Termodin\`amica, Universitat de Val\`encia, \\ E-46100 Burjassot, Spain
\\ garmovla@uv.es
\end{center}
\small{}
\noindent A new class of deterministic dynamical systems, termed semipredictable dynamical systems, is presented. The spatiotemporal evolution of these systems have both predictable and unpredictable traits, as found in natural complex systems. We prove a general result: The dynamics of any deterministic nonlinear cellular automaton (CA) with $p$ possible dynamical states can be decomposed at each instant of time in a superposition of $N$ layers involving $p_{0}$, $p_{1}$,... $p_{N-1}$ dynamical states each, where the $p_{k\in \mathbb{N}}$, $k \in [0, N-1]$ are divisors of $p$. If the divisors coincide with the prime factors of $p$ this decomposition is \emph{unique}. Conversely, we also prove that $N$ CA working on symbols $p_{0}$, $p_{1}$,... $p_{N-1}$ can be composed to create a graded CA rule with $N$ different layers. We then show that, even when the full spatiotemporal evolution can be unpredictable, certain traits (layers) can exactly be predicted. We present explicit examples of such systems involving compositions of Wolfram's 256 elementary CA and a more complex CA rule acting on a neighborhood of two sites and 12 symbols and whose rule table corresponds to the smallest Moufang loop $M_{12}(S_{3},2)$. 
~\\

\noindent complexity; predictability; time series; cellular automata; Moufang loops
\pagebreak


\section{Introduction}

Complex natural systems exhibit a coexistence of predictable and unpredictable traits in their dynamical evolution. This feature seems to have been excluded from mathematical investigation, perhaps because of the difficulty to state the problem rigorously and because of its suspected general mathematical intractability \cite{Wolfram,Wolfram2}. In this article we present a pathway to investigate the problem and in doing so discover an infinite family of systems  termed \emph{semipredictable dynamical systems} whose dynamical trajectory can be sharply separated in the ordinary sum of an exactly predictable part and an unpredictable one,  such that both parts can be neatly extracted from the trajectory at any time. We show how a decomposition of the dynamics can be carried out for any dynamical system modeled by a cellular automaton (CA) in order to extract these layers and make predictions. If the system is totally unpredictable, our method leads, in any case, to a reduction of the dynamics to independent layers that can be treated separately. These layers can then be superimposed to get the full solution. In many cases, however, \emph{although the full solution is unpredictable, there are nontrivial evolving properties of it that can be predicted at any time and we can give an explicit solution for them}. 

In this article we say that a dynamical system is \emph{unpredictable} 
if it is both \emph{deterministic} and \emph{autoplectic} in the sense given by Wolfram to this word in \cite{Wolfram7}: Autoplectic systems are \emph{spatially extended systems that starting from a simple (non-random) initial condition are able to generate intrinsic randomness by themselves so that no regular pattern is discernible in their spatiotemporal evolution}. Autoplectic systems include all computationally irreducible systems for which computation needs to be followed over time to know the outcome of the dynamical variables of the system (for such systems no shortcut for the trajectory generally exist). We call a deterministic spatially extended system \emph{predictable} if it is not autoplectic, its spatiotemporal evolution leading to discernible patterns and regular structures that are amenable to mathematical investigation. Most of Wolfram 256 elementary CAs are predictable. Known exceptions are, among others, rules 30, 110, 54 and their class equivalents under global complementation and left-right transformation \cite{Wolfram, Wolfram7, Chua1, VGM1, VGM2, VGM3}. The complex Ginzburg-Landau equation with parameter values set in the Benjamin-Feir unstable regime leads to turbulence even for simple initial conditions \cite{KuramotoBOOK,contemphys,aranson} and, hence, it constitutes, for such parameter values, another example of autoplectic deterministic (i.e. unpredictable) dynamical system. We find here however, that \emph{there also exist a broad class of dynamical systems of a multilayered nature, some layers being autoplectic and some others being predictable (even trivially predictable) and such that all layers propagate and interfere with each other in all locations during the spatiotemporal evolution of the system}. Indeed, we conjecture that many CA (if not most) with a large number $p$ of dynamical states are semipredictable.

CAs constitute an important tool in the study of complex systems \cite{Wolfram,Hedlund,Adamatzky,Wuensche,Ilachinski,Chua1,Kari,VGM1,arxiv2}. They have been satisfactorily used to model physical \cite{Chopard}, chemical \cite{Kier} and biological \cite{DeutschBOOK} systems and often constitute an alternative to continuous models described by partial differential equations \cite{Toffoli}.  Although physical systems in nature frequently involve many microscopic degrees of freedom, a coarse graining of them can also be generally described by CA \cite{Israeli1,Israeli2} and the latter can thus be used to capture the collective behavior of pattern forming systems. Coarse grained models of CA have been shown to lead to predictable patterns \cite{Israeli1,Israeli2} even when the microscopic dynamics is unpredictable. Since coarse graining always implies that dynamical details are lost and these may be crucial in certain applications, we do not perform here any coarse graining and deal with the spatiotemporal evolution exactly. We show that, while keeping with all spatiotemporal information, it is still often possible to make exact predictions of complex systems whose overall spatiotemporal evolution is unpredictable.


Despite involving only a finite number $p$ of dynamical states and the interactions having a finite range $\rho$, already the most elementary CA can already generate the highest possible complexity \cite{Wolfram, VGM3, Chua1}. In \cite{VGM1} we found a universal map for CA which does not depend on any freely adjustable parameter and which allows any deterministic CA with any number of dynamical states to be described. In \cite{VGM2} and \cite{VGM3} we discussed some symmetries of the universal CA map \cite{VGM2} getting insight in the kind of symmetry breaking that leads to complexity of CA evolution \cite{VGM3}.  In this article we further explore these results by making use of a digit function \cite{QUANTUM, CHAOSOLFRAC} that allows us to express the universal map for CA dynamics in a closed compact form \cite{arxiv2}. One major advantage of this alternative formulation is then made clear, since it allows the CA dynamics to be decomposed into simpler layers that can be separately analyzed, the spatiotemporal evolution of the CA being a superposition of these layers. The outline of this article is as follows. In Section \ref{digit} we introduce the digit function and derive some results in which our approach is based. Then in Section \ref{semip} we apply these results to CAs reducing their dynamics on $p$ symbols to a superposition of (generally nonlinear) dynamical layers, each acting on a reduced number $p_{k\in \mathbb{N}}$ of symbols, where the $p_{k}$'s are divisors of $p$. When these divisors are the prime factors of the canonical decomposition of $p$, the decomposition of the original CA rule is also unique. Our main mathematical result is, thus, a general decomposition property, that is contained in Eqs. (\ref{themapend}) and (\ref{thequations}) below. This result leads us to introduce the concept of $p$-\emph{decomposability}, of relevance to most CA rules in rule space, and the associated concept of graded CA rules, levels and layer CAs. Then, we present several examples of the decomposition giving a detailed account of some semipredictable dynamical systems that are neither totally predictable nor totally unpredictable and analyze it through the mathematical methods presented in this paper. Numerical computations confirm our analytical results.

\section{The digit function} \label{digit}

The digit function, for $p \in \mathbb{N}$, $k \in \mathbb{Z}$ and $x \in \mathbb{R}$ is defined as \cite{QUANTUM} 
\begin{equation}
\mathbf{d}_{p}(k,x)=\left \lfloor \frac{x}{p^{k}} \right \rfloor-p\left \lfloor \frac{x}{p^{k+1}} \right \rfloor    \label{cucuAreal}
\end{equation}
and gives the $k$-th digit of the real number $x$ (when it is non-negative) in a positional numeral system in radix $p > 1$. If $p=1$ the digit function satisfies $\mathbf{d}_{1}(k,x)=\mathbf{d}_{1}(0,x)=0$ and it does not relate to a positional numeral system.

With the digit function we can express any real number as \cite{CHAOSOLFRAC}
\begin{equation}
x=\text{sign}(x)\sum_{k=-\infty}^{\lfloor \log_{p}|x| \rfloor} p^{k} \mathbf{d}_{p}(k,|x|) \label{idenreal}
\end{equation}



The digit function is a staircase of $p$ levels taking discrete integer values between $0$ and $p-1$. Each time that $x$ is divisible by $p^{k+1}$ the ascent of the staircase is broken and the level is  set again to zero and a new staircase begins \cite{CHAOSOLFRAC}. For $n$ and $m$ nonnegative integers the digit function satisfies
\begin{eqnarray}
\mathbf{d}_{p}(k,x+np^{k+1}) &=& \mathbf{d}_{p}(k,x) \label{pro1}\\
\mathbf{d}_{p}(k,p^{k}x) &=&\mathbf{d}_{p}(0,x) \label{pro3} \\
\mathbf{d}_{p}(0,\mathbf{d}_{p}(k,x)) &=& \mathbf{d}_{p}(k,x) \label{pro2} \\
\mathbf{d}_{p}(0,n+\mathbf{d}_{p}(0,m)) &=& \mathbf{d}_{p}(0,n+m) \label{pro2b} \\
\mathbf{d}_{p}(0,n\mathbf{d}_{p}(0,m)) &=& \mathbf{d}_{p}(0,nm) \label{pro2c} \\
\mathbf{d}_{np}(k,x)&=&\mathbf{d}_{p}\left(k, \frac{x}{n^{k}} \right)+p\mathbf{d}_{n}\left(k, \frac{x}{p^{k+1}} \right)=\mathbf{d}_{n}\left(k, \frac{x}{p^{k}} \right)+n\mathbf{d}_{p}\left(k, \frac{x}{n^{k+1}} \right)    \label{reldi1} 
\end{eqnarray}

These relationships are all easy to prove. Eq. (\ref{pro1}) an (\ref{pro3}) follow directly from the definition. Eq. (\ref{pro2}) follows by noting that $\mathbf{d}_{p}(k,x)$ is an integer $0 \le \mathbf{d}_{p}(k,x) \le p-1$. Then 
\begin{equation}
\mathbf{d}_{p}(0,\mathbf{d}_{p}(k,x))=\mathbf{d}_{p}(k,x)-p\left \lfloor \frac{\mathbf{d}_{p}(k,x)}{p} \right \rfloor=\mathbf{d}_{p}(k,x)
\end{equation}
since $\left \lfloor \mathbf{d}_{p}(k,x)/p \right \rfloor=0$. Eq. (\ref{pro2b}) is proved by using Euclidean division, since we can always write $m=ap+b$ with $a, b$ integers and $b=\mathbf{d}_{p}(0,m)$. Therefore
\begin{equation}
\mathbf{d}_{p}(0,n+\mathbf{d}_{p}(0,m))=\mathbf{d}_{p}(0,n+ap+\mathbf{d}_{p}(0,m))=\mathbf{d}_{p}(0,n+m)  
\end{equation}
where Eq. (\ref{pro1}) has also been used. Eq. (\ref{pro2c}) is also proved in a similar way, using the distributive property of ordinary addition and multiplication as well. By using the definition, we have
\begin{eqnarray}
\mathbf{d}_{p}\left(k, \frac{x}{n^{k}} \right)+p\mathbf{d}_{n}\left(k, \frac{x}{p^{k+1}} \right)&=&
\left \lfloor \frac{x}{p^{k}n^{k}} \right \rfloor -p \left \lfloor \frac{x}{p^{k+1}n^{k}} \right \rfloor
+p \left \lfloor \frac{x}{p^{k+1}n^{k}} \right \rfloor-np\left \lfloor \frac{x}{p^{k+1}n^{k+1}} \right \rfloor \nonumber \\
&=& \left \lfloor \frac{x}{(np)^{k}} \right \rfloor -np\left \lfloor \frac{x}{(np)^{k+1}} \right \rfloor =\mathbf{d}_{np}(k,x)
\end{eqnarray}
which proves Eq. (\ref{reldi1}). 

The following result is the cornerstone of our method to decompose CA dynamics as explained in the next section. Let $p_{0}$, $p_{1}$,... $p_{N-1} \in \mathbb{N}$. We have
\begin{equation}
\mathbf{d}_{p_{0}p_{1}\ldots p_{N-1}}\left(k, x \right)=\sum_{h=0}^{N-1}\mathbf{d}_{p_{h}}\left(k, \frac{x}{(\prod_{m=0}^{h-1}p_{m})^{k+1}(\prod_{n=h+1}^{N-1}p_{n})^{k}}\right)\prod_{m=0}^{h-1}p_{m} \label{numeralgen}
\end{equation}
This result can easily be proved by induction. For $N=1$, Eq. (\ref{numeralgen}) is trivially valid and for $N=2$ it reduces to Eq. (\ref{reldi1}). Let us assume the result valid for $N$ factors. Then, for $N+1$ factors we have, by using Eq. (\ref{reldi1})
\begin{eqnarray}
&&\mathbf{d}_{p_{0}p_{1}\ldots p_{N}}\left(k, x \right)=\mathbf{d}_{p_{0}p_{1}\ldots p_{N-1}}\left(k, \frac{x}{p_{N}^{k}} \right)+p_{0}p_{1}\ldots p_{N-1}\mathbf{d}_{p_{N}}\left(k, \frac{x}{(p_{0}p_{1}\ldots p_{N-1})^{k+1}} \right) \nonumber \\
&&=\sum_{h=0}^{N-1}\mathbf{d}_{p_{h}}\left(k, \frac{x/p_{N}^{k}}{(\prod_{m=0}^{h-1}p_{m})^{k+1}(\prod_{n=h+1}^{N-1}p_{n})^{k}}\right)\prod_{m=0}^{h-1}p_{m}+\mathbf{d}_{p_{N}}\left(k, \frac{x}{(\prod_{m=0}^{N-1}p_{m})^{k+1}} \right)\prod_{m=0}^{N-1}p_{m} \nonumber \\
&&=\sum_{h=0}^{N}\mathbf{d}_{p_{h}}\left(k, \frac{x}{(\prod_{m=0}^{h-1}p_{m})^{k+1}(\prod_{n=h+1}^{N}p_{n})^{k}}\right)\prod_{m=0}^{h-1}p_{m}  \nonumber
\end{eqnarray}
which proves the validity of Eq. (\ref{numeralgen}).

\section{The universal CA map and the decomposition of CA dynamics: graded CA rules and semipredictable dynamical systems} \label{semip}

\begin{figure*} 
\includegraphics[width=0.75 \textwidth]{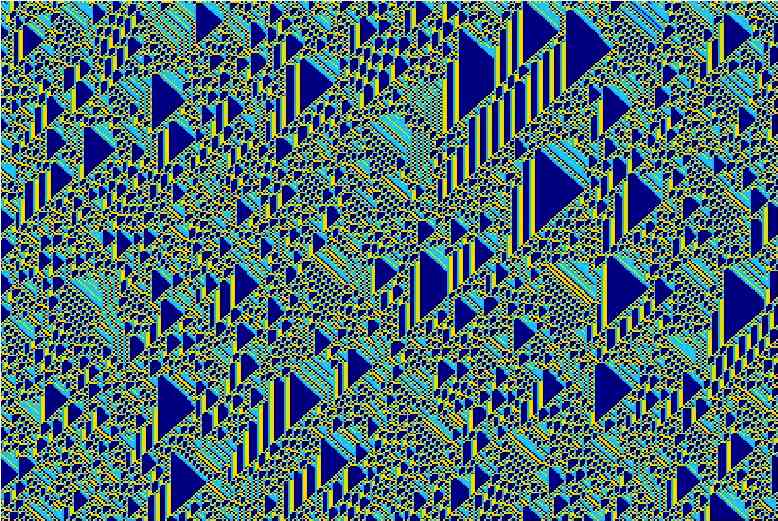}
\caption{\scriptsize{Spatiotemporal evolution of the rule $x_{t+1}^{j}=\mathbf{d}_{3}\left(\sum_{k=-1}^{1}3^{k+r}x_{t}^{j+k} , 741831466113 \right)$ starting from an arbitrary initial condition on a ring of 400 sites and 300 time steps. Time flows from top to bottom. Periodic boundary conditions are used.}} \label{rula}
\end{figure*}

We now establish the way in which the digit function can be used to describe CA. 
Let us consider a 1D ring containing a total number of $N_{s}$ sites. An input is given as initial condition in the form of a vector $\mathbf{x}_{0}=(x_{0}^{1},...,x_{0}^{N_{s}})$. Each of the $x^{j}_{0}$ is an integer in $[0,p-1]$ where superindex $j \in [1, N_{s}]$ specifies the position of the site on the 1D ring. At each $t$ the vector $\mathbf{x}_{t}=(x_{t}^{1},...,x_{t}^{N_{s}})$ specifies the state of the CA.  Periodic boundary conditions are considered so that $x_{t}^{N_{s}+1}=x_{t}^{1}$ and $x_{t}^{0}=x_{t}^{N_{s}}$. Let $x_{t+1}^{j}$ be taken to denote the value of site $j$ at time step $t+1$.  Formally, its dependence on the values at the previous time step is given through the mapping $x_{t+1}^{j}=\ ^{l}R_{p}^{r}(x_{t}^{j+l},\ ... x_{t}^{j}, \ ...,x_{t}^{j-r} )$, which we abbreviate as $x_{t+1}^{j}=\ ^{l}R_{p}^{r}(x_{t}^{j})$ \cite{VGM1} with the understanding that the function on the r.h.s depends on all site values within the neighborhood, with range $\rho=l+r+1$, which contains the site $j$ updated at the next time ($l$ and $r$ denote the number of cells to the left and to the right of site $j$ respectively). We take the convention that $j$ increases to the left. The integer number $n$ in base 10, which runs between $0$ and $p^{r+l+1}-1$, indexes all possible neighborhood values coming from the different configurations of site values. Each of these configurations compares to the dynamical configuration reached by site $j$ and its $r$ and $l$ first-neighbors at time $t$ and given by 
\begin{equation}
n_{t}^{j}=\sum_{k=-r}^{l}p^{k+r}x_{t}^{j+k} \label{NV}
\end{equation}
We will refer to this latter quantity often as the \emph{neighborhood value}. The possible outputs $a_{n}$ for each configuration $n$ are also integers $\in[0,p-1]$. An integer number $R$ can then be given in the decimal base to fully specify the rule $^{l}R_{p}^{r} $ as 
\begin{equation}
R \equiv \sum_{n=0}^{p^{r+l+1}-1}a_{n}p^{n}. \label{RWolf}
\end{equation}
This is the so-called Wolfram code of the CA dynamics. We now note that, from this definition 
\begin{equation}
a_{n}=\mathbf{d}_{p}(n,R) \label{coefi}
\end{equation}
denotes the output of the CA when $n_{t}^{j}=n$. Because $n=\sum_{k=-r}^{l}p^{k+r}\mathbf{d}_{p}(k+r,n)$ (i.e. $n$ can be represented in terms of its radix $p$ expansion), this latter equation is similar to
\begin{equation}
a_{n}=\mathbf{d}_{p}\left(\sum_{k=-r}^{l}p^{k+r}\mathbf{d}_{p}(k+r,n), R\right) \label{coefi2}
\end{equation}
Thus, the universal map for CA is simply given by $x_{t+1}^{j}=a_{n_{t}^{j}}$, which, by noting that $x_{t}^{j+k}=\mathbf{d}_{p}(k+r,n_{t}^{j})$ from Eq. (\ref{NV}), is given by
\begin{equation}
x_{t+1}^{j}=\ ^{l}R_{p}^{r}(x_{t}^{j})=\mathbf{d}_{p}\left(\sum_{k=-r}^{l}p^{k+r}x_{t}^{j+k} , R \right)=\left \lfloor \frac{R}{p^{\sum_{k=-r}^{l}p^{k+r}x_{t}^{j+k}}} \right \rfloor-p \left \lfloor \frac{R}{p^{1+\sum_{k=-r}^{l}p^{k+r}x_{t}^{j+k}}} \right \rfloor 
\label{themap}
\end{equation}
\emph{All deterministic 1D first-order-in-time CA are given by Eq. (\ref{themap}) for any neighborhood range and number of symbols.} With this advantageous compact form for the universal map \emph{we do not need to give any table of configurations as input for the CA but just only the Wolfram code (a nonnegative integer $R \in [0, p^{p^{l+r+1}}])$ and the neighborhood parameters $p$, $l$ and $r$ directly}. For example, the celebrated Wolfram's rule 110 takes the simple form
\begin{equation}
x_{t+1}^{j}=\mathbf{d}_{2}\left(\sum_{k=-1}^{1}2^{k+r}x_{t}^{j+k} , 110 \right)
\end{equation}
The maps for all Wolfram's 256 CA with $l=r=1$ and $p=2$ are thus simply obtained as
\begin{equation}
x_{t+1}^{j}=\mathbf{d}_{2}\left(\sum_{k=-1}^{1}2^{k+r}x_{t}^{j+k} , R \right) \label{WCA}
\end{equation}
by giving an input $R \in [0,255]$. As a further illustration, the `three-color' $p=3$ CA rule $x_{t+1}^{j}=\mathbf{d}_{3}\left(\sum_{k=-1}^{1}3^{k+r}x_{t}^{j+k} , 741831466113 \right)$ is plotted in Fig. \ref{rula} starting from an arbitrary initial condition.


We can now take profit of the results derived in the previous Section for the digit function. Let the number of symbols $p$ have unique prime decomposition into factors $p_{0}$,... ,$p_{N-1}$ as $p=p_{0}p_{1}\ldots p_{N-1}=\prod_{h=0}^{N-1}p_{h}$. Then, by using Eq. (\ref{numeralgen}) in Eq. (\ref{themap}) we obtain
\begin{eqnarray}
x_{t+1}^{j}&=&\sum_{h=0}^{N-1}\mathbf{d}_{p_{h}}\left(\sum_{k=-r}^{l}p^{k+r}x_{t}^{j+k}, \frac{R}{(\prod_{m=0}^{h-1}p_{m})^{1+\sum_{k=-r}^{l}p^{k+r}x_{t}^{j+k}}(\prod_{n=h+1}^{N-1}p_{n})^{\sum_{k=-r}^{l}p^{k+r}x_{t}^{j+k}}}\right)\prod_{m=0}^{h-1}p_{m} \nonumber \\
&=&\sum_{h=0}^{N-1}\mathbf{d}_{p_{h}}\left(\sum_{k=-r}^{l}p^{k+r}x_{t}^{j+k}, \frac{R\ p_{h}^{\sum_{k=-r}^{l}p^{k+r}x_{t}^{j+k}}}{(\prod_{m=0}^{h-1}p_{m})^{1+\sum_{k=-r}^{l}p^{k+r}x_{t}^{j+k}}(p_{h}\prod_{n=h+1}^{N-1}p_{n})^{\sum_{k=-r}^{l}p^{k+r}x_{t}^{j+k}}}\right)\prod_{m=0}^{h-1}p_{m} \nonumber 
\end{eqnarray}
\begin{eqnarray}
&=&\sum_{h=0}^{N-1}\mathbf{d}_{p_{h}}\left(\sum_{k=-r}^{l}p^{k+r}x_{t}^{j+k}, \frac{R\ p_{h}^{\sum_{k=-r}^{l}p^{k+r}x_{t}^{j+k}}}{(\prod_{n=0}^{N-1}p_{n})^{\sum_{k=-r}^{l}p^{k+r}x_{t}^{j+k}}\prod_{m=0}^{h-1}p_{m}}\right)\prod_{m=0}^{h-1}p_{m} \nonumber \\
&=&\sum_{h=0}^{N-1}\mathbf{d}_{p_{h}}\left(\sum_{k=-r}^{l}p^{k+r}x_{t}^{j+k}, \frac{R\ p_{h}^{\sum_{k=-r}^{l}p^{k+r}x_{t}^{j+k}}}{p^{\sum_{k=-r}^{l}p^{k+r}x_{t}^{j+k}}\prod_{m=0}^{h-1}p_{m}}\right)\prod_{m=0}^{h-1}p_{m}  \label{themap2}
\end{eqnarray}
which, by means of Eq. (\ref{pro3}) reduces to
\begin{equation}
x_{t+1}^{j}=\ ^{l}R_{p}^{r}(x_{t}^{j})=\sum_{h=0}^{N-1}\mathbf{d}_{p_{h}}\left(0, \frac{R}{p^{\sum_{k=-r}^{l}p^{k+r}x_{t}^{j+k}}\prod_{m=0}^{h-1}p_{m}}\right)\prod_{m=0}^{h-1}p_{m} \
\label{themapend}
\end{equation}
We say that a CA rule $^{l}R_{p}^{r}$ is \emph{p-decomposable} if $p$ is composite, i.e. not a prime number. If a rule is $p$-decomposable, its decomposition is given by Eq. (\ref{themapend}). Note that if $p$ is prime Eq. (\ref{themapend}) reduces to just only one term in the sum and nothing is gained. 

We observe that Eq. (\ref{themapend}) constitutes a decomposition of the original CA on contributions coming from $N$ digit functions, which represent a spatiotemporal dynamics (layer) acting on $p_{h}$ symbols each. This decomposition is thus produced by adding together all $N$ layers, each of them being accompanied by a factor $\prod_{m=0}^{h-1}p_{m}$. Such weights are responsible for the separation of the original CA into $N$ layers. The latter can display complex behavior since they are generally coupled in a nonlinear fashion and the value on a specific layer generally depends in a complex way on the inputs from other layers. In many cases, however, the spatiotemporal behavior of certain layers can be exactly predicted, as we shall discuss in the next section through examples. First, we note that, for two given integers $a > 1$ and $b$, the digit function satisfies
$\mathbf{d}_{a}(0,b)=0$ if and only if $a$ divides $b$. This last equation is traditionally expressed $b =0 \mod{a}$. Now, we observe that, indeed, from Eq. (\ref{themapend}) (since $0\le \mathbf{d}_{p_{h}}(0,y) \le p_{h}-1$ regardless of the value of the quantity $y$)
\begin{eqnarray}
\mathbf{d}_{p_{0}}(0,x_{t+1}^{j})&=&\mathbf{d}_{p_{0}}(0,\ ^{l}R^{r}_{p}(x_{t}^{j}))=\mathbf{d}_{p_{0}}\left(0, \frac{R}{p^{\sum_{k=-r}^{l}p^{k+r}x_{t}^{j+k}}}\right) \nonumber \\
\mathbf{d}_{p_{0}p_{1}}(0,x_{t+1}^{j})&=&\mathbf{d}_{p_{0}p_{1}}(0,\ ^{l}R^{r}_{p}(x_{t}^{j}))=\mathbf{d}_{p_{0}}\left(0, \frac{R}{p^{\sum_{k=-r}^{l}p^{k+r}x_{t}^{j+k}}}\right)+ \nonumber \\
&&\qquad \qquad \qquad \qquad \quad +\ p_{0}\mathbf{d}_{p_{1}}\left(0, \frac{R}{p^{\sum_{k=-r}^{l}p^{k+r}x_{t}^{j+k}}p_{0}}\right) \nonumber \\
\mathbf{d}_{p_{0}p_{1}p_{2}}(0,x_{t+1}^{j})&=&\mathbf{d}_{p_{0}p_{1}p_{2}}(0,\ ^{l}R^{r}_{p}(x_{t}^{j}))=\mathbf{d}_{p_{0}}\left(0, \frac{R}{p^{\sum_{k=-r}^{l}p^{k+r}x_{t}^{j+k}}}\right)+ \nonumber \\
&&\qquad \qquad \qquad \qquad \quad +\ p_{0}\mathbf{d}_{p_{1}}\left(0, \frac{R}{p^{\sum_{k=-r}^{l}p^{k+r}x_{t}^{j+k}}p_{0}}\right) \nonumber \\
&&\qquad \qquad \qquad \qquad \quad +\ p_{0}p_{1}\mathbf{d}_{p_{2}}\left(0, \frac{R}{p^{\sum_{k=-r}^{l}p^{k+r}x_{t}^{j+k}}p_{0}p_{1}}\right) \nonumber \\
&\ldots& \text{etc.} \nonumber
\end{eqnarray}
And this system of equations, no matter how many layers we have in the composition, can always be  solved for the layers as
\begin{eqnarray}
\mathbf{d}_{p_{0}}\left(0, \frac{R}{p^{\sum_{k=-r}^{l}p^{k+r}x_{t}^{j+k}}}\right)&=& \mathbf{d}_{p_{0}}(0,\ ^{l}R^{r}_{p}(x_{t}^{j}))  \nonumber \\
\mathbf{d}_{p_{1}}\left(0, \frac{R}{p^{\sum_{k=-r}^{l}p^{k+r}x_{t}^{j+k}}p_{0}}\right)&=& \frac{\mathbf{d}_{p_{0}p_{1}}(0,\ ^{l}R^{r}_{p}(x_{t}^{j}))-\mathbf{d}_{p_{0}}(0,\ ^{l}R^{r}_{p}(x_{t}^{j}))}{p_{0}}=\mathbf{d}_{p_{1}}\left(0,\ \frac{^{l}R^{r}_{p}(x_{t}^{j})}{p_{0}}\right) \nonumber \\
\mathbf{d}_{p_{2}}\left(0, \frac{R}{p^{\sum_{k=-r}^{l}p^{k+r}x_{t}^{j+k}}p_{0}p_{1}}\right)&=& \frac{\mathbf{d}_{p_{0}p_{1}p_{2}}(0,\ ^{l}R^{r}_{p}(x_{t}^{j}))-\mathbf{d}_{p_{0}p_{1}}(0,\ ^{l}R^{r}_{p}(x_{t}^{j}))}{p_{0}p_{1}}=\mathbf{d}_{p_{2}}\left(0,\ \frac{^{l}R^{r}_{p}(x_{t}^{j})}{p_{0}p_{1}}\right) \nonumber \\
&\ldots& \text{etc.} \label{decomeggs}
\end{eqnarray}
where Eq. (\ref{reldi1}) has been used in getting to the last identities. We then observe that Eq. (\ref{compota}) can be written as  
\begin{equation}
x_{t+1}^{j}=\ ^{l}R_{p}^{r}(x_{t}^{j})=\sum_{h=0}^{N-1}\mathbf{d}_{p_{h}}\left(0,\ \frac{^{l}R^{r}_{p}(x_{t}^{j})}{\prod_{m=0}^{h-1}p_{m}}\right)\prod_{m=0}^{h-1}p_{m} \label{compotaje}
\end{equation}
which is an alternative form of Eq. (\ref{themapend}) and which, indeed, follows directly from Eq. (\ref{numeralgen}). Thus we have the $N$ equations given by
\begin{equation}
\mathbf{d}_{p_{h}}\left(0,\ \frac{x_{t+1}^{j}}{\prod_{m=0}^{h-1}p_{m}}\right)=\mathbf{d}_{p_{h}}\left(0, \frac{R}{p^{\sum_{k=-r}^{l}p^{k+r}x_{t}^{j+k}}\prod_{m=0}^{h-1}p_{m}}\right) \qquad h=0, 1,\ \ldots, N-1 \label{thequations0}
\end{equation}
which govern the spatiotemporal evolution of each layer $h$ in the composition as a function of the $p$-decomposable rule $^{l}R^{r}_{p}(x_{t}^{j})$ with $p=\prod_{h=0}^{N-1}p_{h}$. We now note that, in general, as a consequence of Eq. (\ref{numeralgen})
\begin{equation}
x_{t}^{j}=\mathbf{d}_{p_{0}}\left(0,x_{t}^{j}\right)+p_{0}\mathbf{d}_{p_{1}}\left(0,\frac{x_{t}^{j}}{p_{0}}\right)+p_{0}p_{1}\mathbf{d}_{p_{2}}\left(0,\frac{x_{t}^{j}}{p_{0}p_{1}}\right)+\ldots=\sum_{h=0}^{N-1}\mathbf{d}_{p_{h}}\left(0,\frac{x_{t}^{j}}{\prod_{m=0}^{h-1}p_{m}}\right)\prod_{m=0}^{h-1}p_{m} \label{cucucompose}
\end{equation}
at every $j$ and $t$. Therefore, if we define $y_{t}^{h,j}\equiv \mathbf{d}_{p_{h}}\left(0,\frac{x_{t}^{j}}{\prod_{m=0}^{h-1}p_{m}}\right)$ so that $y_{t}^{h,j}$ takes values only on the subset $[0,p_{h}-1]$, Eq. (\ref{cucucompose}) can be rewritten as
\begin{equation}
x_{t}^{j}=y_{t}^{0,j}+p_{0}y_{t}^{1,j}+p_{0}p_{1}y_{t}^{2,j}+\ldots=\sum_{h=0}^{N-1}y_{t}^{h,j}\prod_{m=0}^{h-1}p_{m} \label{cucucomposel}
\end{equation}
and, therefore, Eqs. (\ref{thequations0}) take the form
\begin{equation}
y_{t+1}^{h,j}=\mathbf{d}_{p_{h}}\left(0, \frac{R}{p^{\sum_{k=-r}^{l}p^{k+r}\sum_{h=0}^{N-1}y_{t}^{h,j+k}\prod_{m=0}^{h-1}p_{m}}\prod_{m=0}^{h-1}p_{m}}\right) \qquad h=0, 1,\ \ldots, N-1 \label{thequations}
\end{equation}
If we compare this latter expression with Eq. (\ref{themap}) (by using also Eq. (\ref{pro3}) there)
\begin{equation}
x_{t+1}^{j}=\mathbf{d}_{p}\left(0, \frac{R}{p^{\sum_{k=-r}^{l}p^{k+r}x_{t}^{j+k} }} \right)
\end{equation}
we observe that the layer $h$ is described by a CA which is independent of the other layers only if the following relationship is satisfied
\begin{equation}
y_{t+1}^{h,j}=\mathbf{d}_{p_{h}}\left(0, \frac{R}{p^{\sum_{k=-r}^{l}p^{k+r}\sum_{h=0}^{N-1}y_{t}^{h,j+k}\prod_{m=0}^{h-1}p_{m}}\prod_{m=0}^{h-1}p_{m}}\right)=
\mathbf{d}_{p_{h}}\left(0, \frac{L^{(h)}}{p_{h}^{\sum_{k=-r}^{l}p_{h}^{k+r}y_{t}^{h,j+k}}}\right)
  \label{thequationsGRAD}
\end{equation}
for some integer $L^{(h)}$ such that $0\le L^{(h)} \le p_{h}^{l+r+1}-1$ and for all possible values of  $y_{t}^{h,j+k}\in [0,p_{h}-1]$. If a similar expression is satisfied for each layer $h$, we say that the rule $^{l}R^{r}_{p}$ is a \emph{graded rule}. In a graded rule all layers are decoupled and described by a CA. Graded rules are thus a subset of the more general $p$-decomposable rules.

\emph{Eqs. (\ref{themapend}) and (\ref{thequations}) are the main result of this article.} To understand the usefulness of these expressions and to better grasp the concept of a graded CA rule it is appropriate to think as well in the opposite process of constructing a $p$-decomposable rule out of the composition of more elementary CA rules. The following text and the examples in the next section should be helpful to comprehend this abstract approach and its associated unfamiliar notation.


Let us construct a graded rule out of more elementary CA. Let thus $^{l}L_{\ \ p_{0}}^{(0)\  r}$ , $^{l}L_{\ \ p_{1}}^{(1)\  r}$ , $^{l}L_{\ \ p_{2}}^{(2)\  r}$ , etc. be CA rules acting on alphabets of $p_{0}$, $p_{1}$, $p_{2},\ \ldots$ symbols each. Our above approach means that we can construct a $p$-decomposable CA rule $^{l}R^{r}_{p}$, with $p=p_{0}p_{1}p_{2}\ldots$ out of these CA rules in the following way 
\begin{equation}
x_{t+1}^{j}=\ ^{l}R_{p}^{r}(x_{t}^{j})=\ ^{l}L_{\ \ p_{0}}^{(0)\  r}\left(\mathbf{d}_{p_{0}}\left(0,x_{t}^{j}\right)\right)+p_{0}\ ^{l}L_{\ \ p_{1}}^{(1)\  r}\left(\mathbf{d}_{p_{1}}\left(0,\frac{x_{t}^{j}}{p_{0}}\right)\right)+p_{0}p_{1}\ ^{l}L_{\ \ p_{2}}^{(2)\  r}\left(\mathbf{d}_{p_{2}}\left(0,\frac{x_{t}^{j}}{p_{0}p_{1}}\right)\right)+ \ldots \label{compota}
\end{equation}
where $x_{t}^{j+k}\ (k\in [-r,l])$ and $x_{t+1}^{j}$ are all $\in [0,p-1]$. Each $^{l}L_{\ \ p_{h}}^{(h)\  r}$ denotes an elementary CA rule acting on $p_{h}$ symbols. All rules constructed through Eq. (\ref{compota}) are \emph{graded CA rules}. In Eq. (\ref{compota}) we say that rule $^{l}L_{\ \ p_{0}}^{(0)\  r}$ is the \emph{ground rule}, that rule $^{l}L_{\ \ p_{1}}^{(1)\  r}$ is \emph{lifted} by $p_{0}$ and, in general, that rule $^{l}L_{\ \ p_{h}}^{(h)\  r}$ is lifted by a factor $p_{0}p_{1}\ldots p_{h-1}=\prod_{m=0}^{h-1}p_{m}$. In general, we call $p_{0}$, $p_{0}p_{1}$, $p_{0}p_{1}p_{2}, \ \ldots $ the \emph{levels} and the CA rule $^{l}L_{\ \ p_{h}}^{(h)\  r}$ on each level the corresponding \emph{layer CA}.
 Thus, we note that, from Eq. (\ref{compota}),  Eqs. (\ref{thequations}) take the form
\begin{equation}
\mathbf{d}_{p_{h}}\left(0,\ \frac{x_{t+1}^{j}}{\prod_{m=0}^{h-1}p_{m}}\right)=\ ^{l}L_{\ \ p_{h}}^{(h)\  r}\left(\mathbf{d}_{p_{h}}\left(0,\frac{x_{t}^{j}}{\prod_{m=0}^{h-1}p_{m}}\right)\right) \qquad h=0, 1,\ \ldots, N-1 \label{gradequations}
\end{equation}
and,  we observe that, for every $j$ and $t$, this latter equation becomes
\begin{equation}
y_{t+1}^{h,j}=\ ^{l}L_{\ \ p_{h}}^{(h)\  r}\left(y_{t}^{h,j}\right) \qquad h=0, 1,\ \ldots, N-1 \label{gradequations2}
\end{equation}
which is simply the equation of a decoupled CA acting on $p_{h}$ symbols. The $y_{t}^{h,j}$'s evolve within the graded rule and are easily extracted from it through their definition $y_{t}^{h,j}\equiv \mathbf{d}_{p_{h}}\left(0,\frac{x_{t}^{j}}{\prod_{m=0}^{h-1}p_{m}}\right)$. Conversely, the dynamical state $x_{t}^{j}$ of the graded rule is constructed out of the $y_{t}^{h,j}$'s through Eq. (\ref{cucucomposel}).


 Each layer CA  $^{l}L_{\ \ p_{h}}^{(h)\  r}$ has Wolfram code $L^{(h)}=\sum_{n_{h}=0}^{p_{h}-1}(p_{h})^{n_{h}}\ l^{(h)}_{n_{h}}$ where $l_{n_{h}}^{(h)}\equiv \mathbf{d}_{p}(n_{h},L^{(h)})$ is an integer $\in [0,p_{h}-1]$. Now, by looking at the derivation of Eq. (\ref{themap}) from Eqs. (\ref{NV}) to (\ref{coefi2}), we observe that, from the construction, the graded CA rule $^{l}R_{p}^{r}$ has Wolfram code
\begin{equation}
R=\sum_{n=0}^{p^{l+r+1}-1} p^{n}a_{n}=\sum_{n=0}^{p^{l+r+1}-1} p^{n}\sum_{h=0}^{N-1}\mathbf{d}_{p_{h}}\left(\sum_{k=-r}^{l}p_{h}^{k+r}\mathbf{d}_{p_{h}}\left(0, \frac{\mathbf{d}_{p_{h}}\left(k+r, n \right)}{\prod_{m=0}^{h-1}p_{m}} \right),L^{(h)}\right)\prod_{m=0}^{h-1}p_{m}  \label{wolfgrad}
\end{equation}
and, thus, its coefficients in the rule table are
\begin{equation}
a_{n}=\mathbf{d}_{p}(n,R)=\sum_{h=0}^{N-1}\mathbf{d}_{p_{h}}\left(\sum_{k=-r}^{l}p_{h}^{k+r}\mathbf{d}_{p_{h}}\left(0, \frac{\mathbf{d}_{p_{h}}\left(k+r, n \right)}{\prod_{m=0}^{h-1}p_{m}} \right),L^{(h)}\right)\prod_{m=0}^{h-1}p_{m}  \label{cocompota}
\end{equation}
for $n=0,\ldots, p^{l+r+1}-1$.

The following is a summary of some observations:
\begin{itemize}
\item Out of arbitrary CA rules $^{l}L_{\ \ p_{0}}^{(0)\  r}$ , $^{l}L_{\ \ p_{1}}^{(1)\  r}$ , $^{l}L_{\ \ p_{2}}^{(2)\  r},\ \cdots$, we can construct a CA rule $^{l}R^{r}_{p_{0}p_{1}p_{2}\ldots}\equiv \ ^{l}R^{r}_{p}$ which is a graded rule (superposition) of the former. This is provided by Eq. (\ref{compota}).
\item From the knowledge of the graded rule $^{l}R^{r}_{p}=\ ^{l}R^{r}_{p_{0}p_{1}p_{2}\ldots}$ alone, the layer CA rules $^{l}L_{\ \ p_{0}}^{(0)\  r}$ , $^{l}L_{\ \ p_{1}}^{(1)\  r}$ , $^{l}L_{\ \ p_{2}}^{(2)\  r},\ \cdots$ can be extracted at every time and every location. This is what Eqs. (\ref{gradequations}), resp. Eq. (\ref{gradequations2}), mean. 
\item If all factors of a graded rule are equal, $p_{0}=\ldots=p_{N-1}$ then, although a permutation of the layer CA rules do not change the layer content of a graded rule, the spatiotemporal evolution of the resulting graded rule is distinct to the original one (before the permutation) for the same initial condition. Thus, $N!$ same-layer-content graded CA rules with different spatiotemporal evolutions (when starting all from the same initial condition) can be constructed (the layer CAs being the same in all cases, but acting on different levels of the corresponding graded CA rule). The levels \emph{break} the permutation symmetry of the layer CAs within the graded rule. Indeed, a graded CA rule with levels $1$, $p_{0}$, $p_{0}^{2}$,$\ldots$, $p_{0}^{N-1}$ has the form of a \emph{Galois resolvent} (see \cite{Edwards}, pp. 35-37 for an  introduction to this concept and to its usefulness in group theory, as well as Galois' memoir in  Appendix 1 of \cite{Edwards}, i.e. the function $V$ on Lemma II). 
\item We have seen that any factors $p_{0}$,$\ldots $, $p_{N-1}$ give rise to a decomposition in levels $1,\ p_{0},\ p_{0}p_{1}, \ldots$, $\prod_{h=0}^{N-2} p_{h}$. Thus, if some factors are repeated, let $\alpha_{j}$ denote the multiplicity of the factor $p_{j}$ and let $M$ denote the number of distinct factors. Let the index $j$ run over these distinct factors. We then find that there are
\begin{equation}
\frac{N!}{\prod_{j=0}^{M-1}\alpha_{j}!}
\end{equation}
different ways of arranging the same factors into levels. This again leads to respective distinct decompositions with same-layer CA content. For example, for $p=30$ and factors $(2,3,5)$  there are $6!$ graded rules (since we have $\alpha_{0}=\alpha_{1}=\alpha_{2}=1$ and hence $M=N=3$) that can be derived by arranging the factors in different levels $(1, p_{0}, p_{0}p_{1})$ as $(1, 2, 6)$, $(1, 2, 10)$, $(1, 3, 6)$, $(1,3,15)$, $(1,5,10)$, $(1,5,15)$. To these levels correspond, respectively, the arrangements of the layer CAs with $(p_{0},p_{1},p_{2})$ symbols given by $(2, 3, 5)$, $(2, 5, 3)$, $(3, 2, 5)$, $(3,5,2)$, $(5,2,3)$ and $(5,3,2)$.
\item Eq. (\ref{compota}) is \emph{not} equivalent to Eq. (\ref{themapend}) but a \emph{specific case} of the latter. We have that a graded rule is always $p$-decomposable but the contrary is not true. Indeed, if $p$ is composite there are $p^{p^{l+r+1}}$ rules that are $p$-decomposable of which only a subset are graded rules: A $p$-decomposable rule is a graded CA rule only when each layer' dynamics, as generally given by Eq. (\ref{thequations}),  can be brought to the simple form of a layer CA given by Eq. (\ref{gradequations2}).
\item Although there is a \emph{unique} decomposition of a $p$-decomposable rule into the prime factors of $p$, there are a number of other possible decompositions in levels which equals the total number of multiplicative partitions \cite{Hughes} of $p$. \emph{The method above is valid to construct any such decompositions}. For example, since $p=30$ has five multiplicative partitions
$2\times 3 \times 5 = 2 \times 15 = 6 \times 5 = 3 \times 10 = 30$ a CA rule with $p=30$ admits five possible decompositions in factors $(2,3,5)$, $(2,15)$, $(5,6)$, $(3,10)$, $(30)$, each describable by Eqs. (\ref{themapend}) and (\ref{thequations}).  If all $N$ factors of $p$ are distinct (i.e. if $p$ is squarefree) the number of possible decompositions of the graded CA rule is given by the Bell number $B_{N}$, which can explicitly be calculated as \cite{Sharp}
\begin{equation}
B_N=\sum_{k=0}^N \left\{ {N \atop k} \right\}= \sum_{k=0}^N \frac{1}{k!}\sum_{j=0}^{k} (-1)^{k-j} \binom{k}{j} j^n. 
\end{equation}
where $\left\{ {N \atop k} \right\}$ is a Stirling number of the second kind.  
\end{itemize}
Compared to a general $p$-decomposable rule, the layer dynamics greatly simplifies in the case of a graded CA rule, since each layer is independent of the others (even when all interfere with each other in the spatiotemporal evolution of the full graded CA rule). Although the layers are all entangled in the graded CA rule, they can be separated with help of the digit function, as established above.

The above conclusions have still the following important implication. If any layer in Eq. (\ref{themapend}), displays a complex and unpredictable behavior the compound CA rule $^{l}R^{r}_{p}$ will also generally display a complex and unpredictable behavior. However, certain layers within a $p$-decomposable rule may be predictable, even when some others are not (and the composition is certainly not). We term such CA rules \emph{semipredictable} to emphasize the fact that shortcuts or even closed forms can sometimes be found for the orbits of some of the separate layers. The examples in the next section substantiate this statement.

\section{Examples and discussion}

\subsection{Graded CA rules from Wolfram's elementary CA}

Let us first consider an example of a construction of a graded CA rule out of two simpler rules. 
Wolfram's rule $^{1}30^{1}_{2}$ is well-known to display an unpredictable, chaotic behavior \cite{Wolfram4}. This rule belongs to Wolfram's 256 elementary CA with two symbols ($p=2$) and a neighborhood with one left and one right neighbors ($l=r=1$). On the contrary, Wolfram's rule $^{1}150^{1}_{2}$ is a predictable (albeit complex and subtle) \cite{Martin} CA displaying a nested pattern. Wolfram's rule $^{1}150^{1}_{2}$ has map $x_{t+1}^{j}=x_{t}^{j+1}+_{2}x_{t}^{j}+_{2}x_{t}^{j-1}$ where $+_{2}$ denotes addition modulo 2 (\cite{VGM2,VGM3}) and, thus, is a linear CA. Such CA are \emph{not} autoplectic \cite{Wolfram7}. The explicit solution for the orbit as a function of the initial condition can indeed be found in some references (see e.g. \cite{Chua1}). Although such solution depends on a sum with a number $t$ of terms taken from the initial condition and weighted with binomial coefficients, fruitful use can be made in most cases of the Lucas' correspondence theorem \cite{VGM3}, to simplify the resulting expression. We shall discuss linear CA in more detail elsewhere in connection with Lucas' theorem. 

We can use Eq. (\ref{compota}) to construct a CA with $p=4$, $p_{0}=p_{1}=2$ out of these two rules as
\begin{eqnarray}
x_{t+1}^{j}&=&\ ^{1}R_{4}^{1}(x_{t}^{j})=\ ^{1}30^{1}_{2}\left(\mathbf{d}_{2}\left(0 , x_{t}^{j} \right)\right)+2\cdot \  ^{1}150^{1}_{2}\left(\mathbf{d}_{2}\left(0 , \frac{x_{t}^{j}}{2} \right)\right) \nonumber \\
&=&\mathbf{d}_{2}\left(\sum_{k=-1}^{1}2^{k+1}\mathbf{d}_{2}\left(0 , x_{t}^{j+k} \right) , 30 \right)+2\mathbf{d}_{2}\left(\sum_{k=-1}^{1}2^{k+1}\mathbf{d}_{2}\left(0 , \frac{x_{t}^{j+k}}{2} \right) , 150 \right)  \label{Wolfe1}
\end{eqnarray}
This is a  graded CA rule acting on four symbols $0,\ 1,\ 2,\ 3$ which are the values that $x_{t+1}^{j}$ can take at each  time $t$ and for every $j$. The Wolfram code of the graded rule is easily calculated from Eq. (\ref{wolfgrad}) as
\begin{eqnarray}
R&=&\sum_{n=0}^{4^{3}-1} 4^{n}a_{n}=\sum_{n=0}^{63} 4^{n}a_{n} \label{Rex1}
\end{eqnarray}
where the $a_{n}$'s are given by Eq. (\ref{cocompota}) as
\begin{equation}
a_{n}=\mathbf{d}_{4}(n,R)=\sum_{h=0}^{1}2^{h}\mathbf{d}_{2}\left(\sum_{k=-1}^{1}2^{k+1}\mathbf{d}_{2}\left(0, \frac{\mathbf{d}_{2}\left(k+1, n \right)}{2^{h}} \right),L^{(h)}\right) \qquad (n=0,\ \ldots, 63) \label{cocompotaex1}
\end{equation}
with $L^{(0)}=30$ and $L^{(1)}=150$. Thus, the rule vector of the graded CA rule is obtained from this expression
\begin{eqnarray}
&&(a_{0}, a_{1}, \ldots, a_{63})= \nonumber \\
&&(0,\ 1,\ 2,\ 3,\ 1,\ 1,\ 3,\ 3,\ 2,\ 3,\ 0,\ 1,\ 3,\ 3,\ 1,\ 1,\ 1,\ 0,\ 3,\ 2,\ 0,\ 0,\ 2,\ 2,\ 3,\ 2,\ 1,\ 0,\ 2,\ 2,\ \nonumber \\ &&0,\ 0,\ 2,\ 3,\ 0,\ 1,\ 3,\ 3,\ 1,\ 1,\ 0,\ 1,\ 2,\ 3,\ 1,\ 1,\ 3,\ 3,\ 3,\ 2,\ 1,\ 0,\ 2,\ 2,\ 0,\ 0,\ 1,\ 0,\ 3,\ 2,\ \nonumber \\
&& 0,\ 0,\ 2,\ 2) \label{ruvecex1}
\end{eqnarray}
The rule has a huge Wolfram code of order $R \sim 4^{63}$. The spatiotemporal evolution of the rule and the layer CAs are dictated by Eqs. (\ref{gradequations2}) and (\ref{cucucomposel}), which take the form
\begin{eqnarray}
y_{t+1}^{0,j}&=&\ ^{1}30_{2}^{1}\left(y_{t}^{0,j}\right)=\mathbf{d}_{2}\left(\sum_{k=-1}^{1}2^{k+r}y_{t}^{j+k} , 30 \right) \qquad \qquad y_{0}^{0,j}=\mathbf{d}_{2}(0,x_{0}^{j})  \label{30lift150a} \\
y_{t+1}^{1,j}&=&\ ^{1}150_{2}^{1}\left(y_{t}^{1,j}\right)=\mathbf{d}_{2}\left(\sum_{k=-1}^{1}2^{k+r}y_{t}^{j+k} , 150 \right) \qquad \quad y_{0}^{1,j}=\mathbf{d}_{2}(0,x_{0}^{j}/2) \label{30lift150b} \\
x_{t+1}^{j}&=&y_{t+1}^{0,j}+2y_{t+1}^{1,j}    \label{30lift150c}
\end{eqnarray} 
starting from an initial condition $x_{0}^{j}$. As a consequence of the main result of this article, these equations yield the full description of the spatiotemporal dynamics of the graded CA rule with 4 symbols given by
\begin{equation}
x_{t+1}^{j}=\ ^{1}R_{4}^{1}(x_{t}^{j})=\mathbf{d}_{4}\left(\sum_{k=-1}^{1}4^{k+1}x_{t}^{j+k} , R \right) \label{morecomplex1}
\end{equation}
with $R$ given by Eqs. (\ref{Rex1}) and (\ref{ruvecex1}).

\begin{figure*} 
\includegraphics[width=1.0 \textwidth]{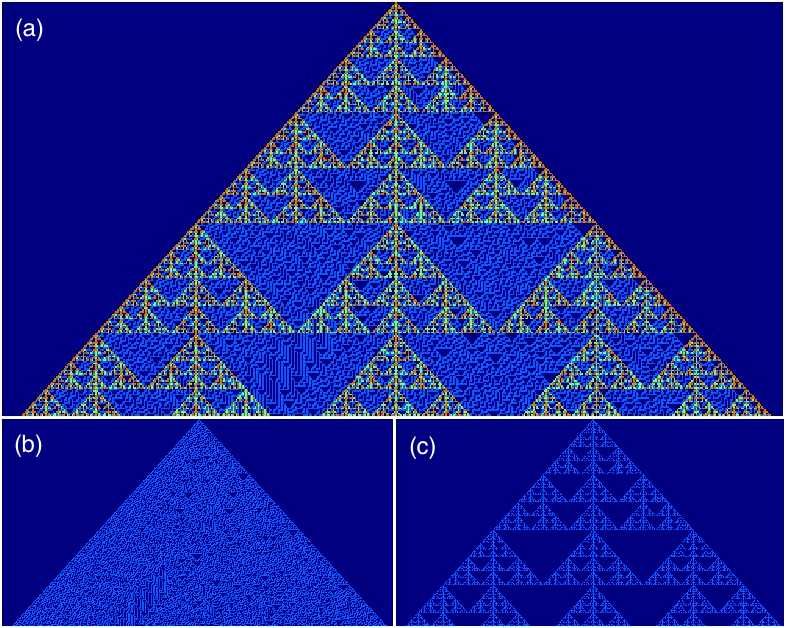}
\caption{\scriptsize{Spatiotemporal evolution of (a) $x_{t}^{j}$, (b) $y_{t}^{0,j}=\mathbf{d}_{2}(0,x_{t}^{j})$ and (c) $y_{t}^{1,j}=\mathbf{d}_{2}(0,x_{t}^{j}/2)$ all calculated by iterating Eq. (\ref{morecomplex1}) (see text), for the CA rule with code given by Eqs. (\ref{Rex1}) and (\ref{ruvecex1}) and for initial conditions corresponding to $x_{0}^{250}=3$ and $x_{0}^{j}=0$ for $j \ne 250$.  We observe that the rule is four-color graded rule, with layer CAs consisting of the two-color Wolfram rules $^{1}30^{1}_{2}$ as ground rule, and $^{1}150^{1}_{2}$, as shown respectively in (b) and (c). Time flows from top to bottom in each panel. Periodic boundary conditions are used, $j \in [1,500]$ and time $t \in [0,240]$.}}  \label{spatiosEX1}
\end{figure*}

We have performed numerical calculations of the spatiotemporal evolution of the CA above. On one hand, it is straightforward to obtain the evolution of the rule from Eqs. (\ref{30lift150a}) to (\ref{30lift150c}). On the other hand, calculating the evolution directly from Eq. (\ref{morecomplex1}) is a bit more demanding because of the large powers of the radix $p=4$ in the digit function. This problem is computationally overcome in the light of our previous work  \cite{VGM1}: We consider the rule vector, as given by Eq. (\ref{ruvecex1}), and use the universal map for CA derived in \cite{VGM1} 
\begin{equation}
x_{t+1}^{j}=\ ^{l}R_{p}^{r}(x_{t}^{j})=\sum_{n=0}^{p^{r+l+1}-1}a_{n}\mathcal{B}\left(n-\sum_{k=-r}^{l}p^{k+r}x_{t}^{j+k}, \frac{1}{2} \right) \label{CA}
\end{equation}
where
\begin{equation}
\mathcal{B}(x,y) \equiv \frac{1}{2}\left(\frac{x+y}{|x+y|}-\frac{x-y}{|x-y|}\right) = \frac{1}{2}\left(\text{sign}(x+y)-\text{sign}(x-y)\right)  \label{d1}
\end{equation} 
is the $\mathcal{B}$-function for any real numbers $x$, $y$. This function returns $\text{sign}(y)$ if $-y<x<y$, $\text{sign}(y)/2$ if $x=y$ and zero otherwise. Eq. (\ref{CA}) is equivalent to Eq. (\ref{themap}) and overcomes the problem of large powers of $p$. While the digit function has mathematical advantages compared to the $\mathcal{B}$-function, the latter is more suitable for intensive computations. In Eq. (\ref{CA}) we thus take $p=4$, $l=1, r=1$ and the $a_{n}$ given by Eq. (\ref{ruvecex1}). In Fig. \ref{spatiosEX1} we show the spatiotemporal evolution of (a) $x_{t}^{j}$, (b) $\mathbf{d}_{2}(0,x_{t}^{j})$, (c) $\mathbf{d}_{2}(0,x_{t}^{j}/2)$ calculated from Eq. (\ref{morecomplex1}) (i.e. Eq. (\ref{CA}) with the rule parameters $a_{n}$ given by Eq. (\ref{ruvecex1})) and for a simple initial condition. We thus see that such a CA rule allows both chaotic and coherent signals to be spatially transmitted \emph{together} in different layers, starting from a simple initial condition \emph{within the same} compound signal. Although both layers interfere in forming the pattern of the graded rule, they can be neatly separated by means of the digit function into two different signals both employing two symbols and behaving quite differently: One of them, albeit complex, has a large degree of coherence; the other one is incoherent. \emph{All these statements are true for any arbitrary initial condition.} We have calculated the graded rule both from Eqs. (\ref{30lift150c}) to (\ref{30lift150c}), and from Eq. (\ref{morecomplex1}) (resp. Eq. (\ref{CA})),  computationally confirming the equivalence of both approaches.

\begin{figure*} 
\includegraphics[width=1.0 \textwidth]{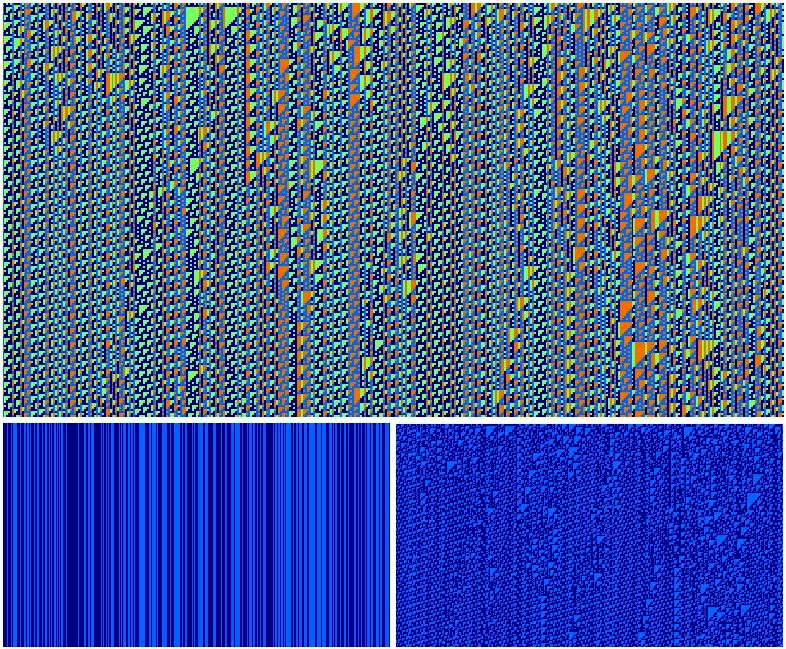}
\caption{\scriptsize{Spatiotemporal evolution of $x_{t}^{j}$ (top), $y_{t}^{0,j}=\mathbf{d}_{2}(0,x_{t}^{j})$ (bottom left) and $y_{t}^{1,j}=\mathbf{d}_{2}(0,x_{t}^{j}/2)$ (bottom right) all calculated by iterating Eq. (\ref{morecomplex1}) (see text), for the CA rule with code given by Eqs. (\ref{Rex1}) and (\ref{ruvecex2}) and for a random initial condition.  The rule is a four-color graded CA rule, with layer CAs consisting of the two-color Wolfram rules $^{1}204^{1}_{2}$ (the identity) as ground rule and $^{1}137^{1}_{2}$ (capable of universal computation \cite{Cook,Wolfram,Chua1}) as lifted rule.  \emph{The graded CA rule in the top panel is thus  a semipredictable system.} Time flows from top to bottom in each panel. Periodic boundary conditions are used, $j \in [1,400]$ and time $t \in [0,200]$.}}  \label{spatiosEX2}
\end{figure*}

Another example is provided by layer CA rules $^{1}204^{1}$ and $^{1}137^{1}$ with Wolfram codes  $L^{(0)}=204$ and $L^{(1)}=137$. The rule vector of the graded rule is calculated from Eq. (\ref{cocompotaex1}) as
\begin{eqnarray}
&&(a_{0}, a_{1}, \ldots, a_{63})= \nonumber \\
&&(2,\ 2,\ 0,\ 0,\ 3,\ 3,\ 1,\ 1,\ 0,\ 0,\ 2,\ 2,\ 1,\ 1,\ 3,\ 3,\ 2,\ 2,\ 0,\ 0,\ 3,\ 3,\ 1,\ 1,\ 0,\ 0,\ 2,\ 2,\ 1,\ 1,\ \nonumber \\
&& 1,\ 3,\ 3,\ 0,\ 0,\ 0,\ 0,\ 1,\ 1,\ 1,\ 1,\ 0,\ 0,\ 2,\ 2,\ 1,\ 1,\ 3,\ 3,\ 0,\ 0,\ 0,\ 0,\ 1,\ 1,\ 1,\ 1,\  0,\ 0,\ 2,\ \nonumber \\
&& 2,\ 1,\ 1,\ 3,\ 3) \label{ruvecex2}
\end{eqnarray}
The spatiotemporal evolution of the graded rule is shown in Fig. \ref{spatiosEX2} together with the ones of its layer CAs. CA rule $^{1}204^{1}$ is trivially predictable since it is equal to the identity and one has $y_{t}^{0,j}=y_{0}^{0,j}$ for every $t$. On the contrary, rule $^{1}137^{1}$ displays complex and unpredictable behavior (the rule is just the global complement of rule $^{1}110^{1}$ which is known to emulate a universal Turing machine \cite{Cook, Wolfram}) and no shortcut for the orbit of $y_{t}^{1,j}$ is possible. Consequently, no shortcut is possible for the graded CA rule, whose evolution is given by $x_{t}^{j}=y_{t}^{0,j}+2y_{t}^{1,j}$ (top panel). The unpredictable character of the graded rule is entirely given by that of the rule $^{1}137_{2}^{1}$. Thus the rule vector in Eq. (\ref{ruvecex2}) yields a graded rule $^{1}R^{1}_{4}$ which is a semipredictable system one layer being trivially predictable and the other one not.

We thus note that all $4^{4^{3}}=4^{256} \sim 3\cdot 10^{38}$ CA rules $^{1}R^{1}_{4}$ are $p$-decomposable. Of these, only a subset consisting of $2^{2^{3}}2^{2^{3}}=256^{2}=65536$ are graded rules, obtained from choosing any two of the 256 elementary Wolfram CA rules as ground and lifted layer CAs. 

\begin{figure*} 
\includegraphics[width=1.0 \textwidth]{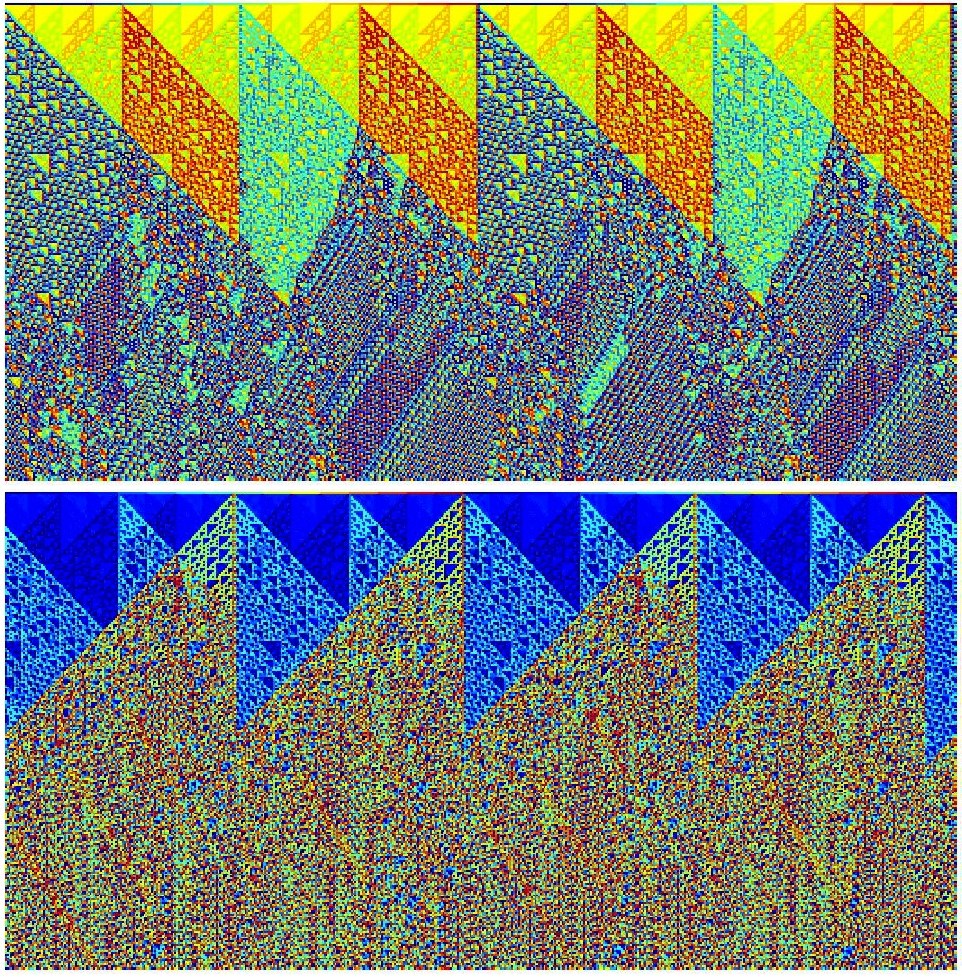}
\caption{\scriptsize{Spatiotemporal evolution of $x_{t}^{j}$ for rule A (see text) given by Eqs. (\ref{110a}) to (\ref{110to193}) (top), and for rule B (bottom) obtained by a permutation of the layers of rule A (see the text). Both rules have number of symbols $p=16$ and neighborhood range $l+r+1=3$. The initial condition is equal in both cases and taken as $x_{0}^{j}=\mathbf{d}_{16}(1,j)$. Thus, although the layer content of both graded rules is the same and the initial condition is also identical in both cases their spatiotemporal evolutions are quite different because graded rules are not invariant under permutation of the layers. Time flows from top to bottom in each panel. Periodic boundary conditions are used, with $j \in [1,528]$ and time $t \in [0,250]$.}}  \label{unirul}
\end{figure*}

We now construct more complex graded CA rules with $p=16$, arising as the result of the composition of the CA rules $^{1}110_{2}^{1}$, $^{1}124_{2}^{1}$, $^{1}137_{2}^{1}$ and $^{1}193_{2}^{1}$. All these CAs are related by global complementation and/or left-right transformation and all four are capable of universal computation. Thus, the corresponding $4!=24$ graded CA rules that can be formed with these four layer CAs are all universal as well. Let us consider the graded CA rule given by
\begin{eqnarray}
y_{t+1}^{0,j}&=&\ ^{1}110_{2}^{1}\left(y_{t}^{0,j}\right)=\mathbf{d}_{2}\left(\sum_{k=-1}^{1}2^{k+r}y_{t}^{0,j+k} , 110 \right) \qquad \quad y_{0}^{0,j}=\mathbf{d}_{2}(0,x_{0}^{j})  \label{110a} \\
y_{t+1}^{1,j}&=&\ ^{1}124_{2}^{1}\left(y_{t}^{1,j}\right)=\mathbf{d}_{2}\left(\sum_{k=-1}^{1}2^{k+r}y_{t}^{1,j+k} , 124 \right) \qquad \quad y_{0}^{1,j}=\mathbf{d}_{2}(0,x_{0}^{j}/2) \label{124b} \\
y_{t+1}^{2,j}&=&\ ^{1}137_{2}^{1}\left(y_{t}^{2,j}\right)=\mathbf{d}_{2}\left(\sum_{k=-1}^{1}2^{k+r}y_{t}^{2,j+k} , 137 \right) \qquad \quad y_{0}^{2,j}=\mathbf{d}_{2}(0,x_{0}^{j}/4) \label{137c} \\
y_{t+1}^{3,j}&=&\ ^{1}193_{2}^{1}\left(y_{t}^{3,j}\right)=\mathbf{d}_{2}\left(\sum_{k=-1}^{1}2^{k+r}y_{t}^{3,j+k} , 193 \right) \qquad \quad y_{0}^{3,j}=\mathbf{d}_{2}(0,x_{0}^{j}/8) \label{193d} \\
x_{t+1}^{j}&=&y_{t+1}^{0,j}+2y_{t+1}^{1,j}+4y_{t+1}^{2,j}+8y_{t+1}^{3,j}    \label{110to193}
\end{eqnarray} 
The above system of equations completely specify the spatiotemporal evolution of the rule. Let us call it rule A. Now, if we e.g. make the transformation $h \to 3-h$ in $y_{t}^{h,j}$, $\forall j$ and $t$, the result is another graded CA rule with the same layer content than the original one, but leading to a different spatiotemporal evolution for the same initial condition. This is so because of the symmetry breaking introduced by the different level weights 1, 2, 4 and 8. Let us call this rule so obtained rule B. In Fig.\ref{unirul} the spatiotemporal evolutions of rule $A$ (top) and rule $B$ (bottom) are plotted for an initial condition given by $x_{0}^{j}=\mathbf{d}_{16}(1,j)$. Although the decomposition in layers of both rules is the same, as is the initial condition, the spatiotemporal evolution looks very different. Rule A displays a proliferation of coherent structures creating a kind of granular texture. Rule B seems to lead to more disorder. 



\subsection{A non-graded $p$-decomposable rule with semipredictable dynamics}  

In the previous examples we have considered the construction of $p$-decomposable graded rules from layer CA rules. In the following example we consider the opposite process: Given a $p$-decomposable CA rule, we derive and analyze its layers and attempt to reach conclusions on its dynamics. We consider a $p$-decomposable rule that, as we shall see is not a graded rule.

Let us consider a rule $^{0}R^{1}_{12}$, i.e. with $p=12$, $l=0$, $r=1$, range $\rho=l+r+1=2$,  Wolfram code
\begin{equation}
R=\sum_{n=0}^{143}a_{n}12^{n} \label{mouCA}
\end{equation}
and rule vector 
\begin{eqnarray}
&&(a_{0},\ a_{1},\ldots,\ a_{143})= (0,\ 1,\ 2,\ 3,\ 4,\ 5,\ 6,\ 7,\ 8,\ 9,\ 10,\ 11,\ 1,\ 0,\ 4,\ 5,\ 2,\ 3,\ 7,\ 6,\ 10,\ \nonumber \\ && 11,\ 8,\ 9,\ 2,\ 5,\ 0,\ 4,\ 3,\ 1,\ 8,\ 11,\ 6,\ 10,\ 9,\ 7,\ 3,\ 4,\ 5,\ 0,\ 1,\ 2,\ 9,\ 10,\ 11,\ 6,\ 7,\ 8,\ 4,\  \nonumber \\&&  3,\ 1,\ 2,\ 5,\ 0,\ 10,\ 9,\ 7,\ 8,\ 11,\ 6,\ 5,\ 2,\ 3,\ 1,\ 0,\ 4,\ 11,\ 8,\ 9,\ 7,\ 6,\ 10,\ 6,\ 7,\ 8,\ 9,\ 11,\  \nonumber \\ && 10,\ 0,\ 1,\ 2,\ 3,\ 5,\ 4,\ 7,\ 6,\ 10,\ 11,\ 9,\ 8,\ 1,\ 0,\ 5,\ 4,\ 2,\ 3,\ 8,\ 11,\ 6,\ 10,\ 7,\ 9,\ 2,\ 4,\ 0,\  \nonumber \\&& 5,\ 3,\ 1,\ 9,\ 10,\ 11,\ 6,\ 8,\ 7,\ 3,\ 5,\ 4,\ 0,\ 1,\ 2,\ 10,\ 9,\ 7,\ 8,\ 6,\ 11,\ 4,\ 2,\ 3,\ 1,\ 0,\ 5,\ 11,\  \nonumber \\&& 8,\ 9,\  7,\ 10,\ 6,\ 5,\ 3,\ 1,\ 2,\ 4,\ 0) \label{mouCA2}
\end{eqnarray}
This rule provides the map $x_{t+1}^{j}=\ ^{0}R^{1}_{12}(x_{t}^{j}) \equiv x_{t}^{j}*x_{t}^{j-1}$ which is described by Eq. (\ref{themap}) and which shows that the CA rule $^{0}R^{1}_{12}$ above can be understood as a binary operator $*$ (law of composition) $S\times S \to S$ where $S$ is the set of the integers $\in [0,11]$. We note that, $n=\mathbf{d}_{12}\left(0,n\right)+12\mathbf{d}_{12}\left(1,n\right)$ ($n \in [0,144]$) and that $n_{t}^{j}=12x_{t}^{j}+x_{t}^{j-1}$ where $x_{t}^{j}= \mathbf{d}_{12}(1,n_{t}^{j})$ and $x_{t}^{j-1}= \mathbf{d}_{12}(0,n_{t}^{j})$. Thus, since $x_{t+1}^{j}=a_{n_{t}^{j}}=x_{t}^{j}*x_{t}^{j-1}=\mathbf{d}_{12}\left(1,n_{t}^{j}\right)*\mathbf{d}_{12}\left(0,n_{t}^{j}\right)$, it proves useful to arrange the rule vector on a table so that the output $a_{n}=\mathbf{d}_{12}\left(1,n\right)*\mathbf{d}_{12}\left(0,n\right)$ is obtained from reading the table for any $\mathbf{d}_{12}\left(1,n\right)$ listed in the rows and and $\mathbf{d}_{12}\left(0,n\right)$ listed in the columns of the table. The latter has the form
\begin{center}
\begin{tabular}{c|cccccccccccc}
 $\ * \ $ & $\ 0 \ $ & $\ 1 \ $ & $\ 2 \ $ & $\ 3\ $ & $ \ 4 \ $ &$\ 5 \ $&$\ 6 \ $&$\ 7 \ $&$\ 8 \ $&$\ 9 \ $&$\ 10 \ $&$\ 11 \ $ \\
\hline
$\ 0 \ $ & $\ 0 \ $ & $\ 1 \ $ & $\ 2 \ $ & $\ 3\ $ & $ \ 4 \ $ &$\ 5 \ $ &$\ 6 \ $&$\ 7 \ $&$\ 8 \ $&$\ 9 \ $&$\ 10 \ $&$\ 11 \ $ \\
$\ 1 \ $ & $\ 1 \ $ & $\ 0 \ $ & $\ 4 \ $ & $\ 5\ $ & $\ 2 \ $ &$\ 3 \ $&$ \ 7 \ $&$ \ 6 \ $&$\ 10 \ $&$\ 11 \ $&$\ 8 \ $&$\ 9 \ $ \\
$\ 2 \ $ & $\ 2 \ $ & $\ 5 \ $ & $\ 0 \ $ & $\ 4\ $&$\ 3 \ $&$\ 1 \ $&$\ 8 \ $&$\ 11 \ $&$\ 6 \ $&$\ 10 \ $&$\ 9 \ $&$\ 7 \ $ \\
$\ 3 \ $ & $\ 3 \ $ & $\ 4 \ $ & $\ 5 \ $ & $\ 0\ $&$\ 1 \ $&$\ 2 \ $&$\ 9 \ $&$\ 10 \ $&$\ 11 \ $&$\ 6 \ $&$\ 7 \ $&$\ 8 \ $ \\
$\ 4 \ $ & $\ 4 \ $ & $\ 3 \ $ & $\ 1 \ $ & $\ 2\ $&$\ 5 \ $&$\ 0 \ $&$\ 10 \ $&$\ 9 \ $&$\ 7 \ $&$\ 8 \ $&$\ 11 \ $&$\ 6 \ $ \\
$\ 5 \ $ & $\ 5 \ $ & $\ 2 \ $ & $\ 3 \ $ & $\ 1 \ $&$\ 0 \ $&$\ 4 \ $&$\ 11 \ $&$\ 8 \ $&$\ 9 \ $&$\ 7 \ $&$\ 6 \ $&$\ 10 \ $ \\
$\ 6 \ $ & $\ 6 \ $&$\ 7 \ $&$\ 8 \ $&$\ 9 \ $&$\ 11 \ $&$\ 10 \ $&$\ 0 \ $ & $\ 1 \ $ & $\ 2 \ $ & $\ 3\ $ & $ \ 5 \ $ &$\ 4 \ $ \\
$\ 7 \ $ & $ \ 7 \ $&$ \ 6 \ $&$\ 10 \ $&$\ 11 \ $&$\ 9 \ $&$\ 8 \ $&$\ 1 \ $ & $\ 0 \ $ & $\ 5 \ $ & $\ 4\ $ & $\ 2 \ $ &$\ 3 \ $ \\
$\ 8 \ $ & $\ 8 \ $&$\ 11 \ $&$\ 6 \ $&$\ 10 \ $&$\ 7 \ $&$\ 9 \ $&$\ 2 \ $ & $\ 4 \ $ & $\ 0 \ $ & $\ 5\ $&$\ 3 \ $&$\ 1 \ $ \\
$\ 9 \ $ & $\ 9 \ $&$\ 10 \ $&$\ 11 \ $&$\ 6 \ $&$\ 8 \ $&$\ 7\ $ & $\ 3 \ $ & $\ 5 \ $ & $\ 4 \ $ & $\ 0\ $&$\ 1 \ $&$\ 2 \ $ \\
$\ 10 \ $ & $\ 10 \ $&$\ 9 \ $&$\ 7 \ $&$\ 8 \ $&$\ 6 \ $&$\ 11 \ $ &$\ 4 \ $ & $\ 2 \ $ & $\ 3 \ $ & $\ 1\ $&$\ 0 \ $&$\ 5 \ $ \\
$\ 11 \ $ & $\ 11 \ $&$\ 8 \ $&$\ 9 \ $&$\ 7 \ $&$\ 10 \ $&$\ 6 \ $& $\ 5 \ $ & $\ 3 \ $ & $\ 1 \ $ & $\ 2 \ $&$\ 4 \ $&$\ 0 \ $ \\
\end{tabular}
\end{center} 

This table corresponds to the Cayley table of the smallest Moufang loop $M_{12}(S_{3},2)$ \cite{Chein2} of order 12 (see \cite{Chein} for a classification of small Moufang loops). A law of composition $*$, $S\times S \to S$  is a Moufang loop if it satisfies the following identities, called the Moufang identities \cite{Moufang} 
\begin{eqnarray}
    z*(x*(z*y)) &=& ((z*x)*z)*y \\
    x*(z*(y*z)) &=& ((x*z)*y)*z \\
    (z*x)*(y*z) &=& (z*(x*y))*z \\
    (z*x)*(y*z) &=& z*((x*y)*z)
\end{eqnarray}
for $x,y,z \in S$. It can be checked from the table above that this is indeed the case. Here $S$ is the set of integers $\in [0,11]$. We see from the table above and the Moufang identities that the operator $*$ is both non-commutative and non-associative.
However, if $x=z$, $x=y$ or $y=z$, the resulting subalgebra is associative. We thus expect that (as a consequence of losing the associative property) a high degree of disorder/unpredictability in the general spatiotemporal dynamics and (as a consequence of the existence of associative subalgebras) the existence of some regimes where the dynamics is simpler to describe.

Let us investigate the above facts with the mathematical methods presented in the previous section. We have that $p=12=2^{2}\cdot 3$ is composite and admits multiplicative partitions with factors $(p_{0},p_{1},p_{2},p_{3})$ given by $(2,2,3)$, $(4,3)$, $(6,2)$ and the trivial (non-decomposed) one $(12)$. We study the partition $(6,2)$ with the layer $p_{0}=6$ as ground layer and the layer with $p_{1}=2$ lifted by 6. From Eq.(\ref{cucucomposel}) and Eqs. (\ref{thequations}) we have that the dynamics of the layers and of the $p$-decomposable rule are given by
\begin{eqnarray}
y_{t+1}^{0,j}&=&\mathbf{d}_{6}\left(0, \frac{\sum_{n=0}^{143}a_{n}12^{n}}{12^{\sum_{k=-1}^{0}12^{k+1}\sum_{h=0}^{1}y_{t}^{h,j+k}\prod_{m=0}^{h-1}p_{m}}}\right) \qquad \ \qquad y_{0}^{0,j}=\mathbf{d}_{6}(0,x_{0}^{j}) \label{thequaMou1} \\
y_{t+1}^{1,j}&=&\mathbf{d}_{2}\left(0, \frac{\sum_{n=0}^{143}a_{n}12^{n}}{6\cdot 12^{\sum_{k=-1}^{0}12^{k+1}\sum_{h=0}^{1}y_{t}^{h,j+k}\prod_{m=0}^{h-1}p_{m}}}\right)  \qquad \quad y_{0}^{1,j}=\mathbf{d}_{2}(0,x_{0}^{j}/6) \label{thequaMou2}\\
x_{t+1}^{j}&=&y_{t+1}^{0,j}+6y_{t+1}^{1,j}    \label{thequaMou3}
\end{eqnarray}
for the $a_{n}$'s given by Eq. (\ref{mouCA2}). After a little algebra, these equations reduce to
\begin{eqnarray}
y_{t+1}^{0,j}&=&\mathbf{d}_{6}\left(0, \sum_{n=0}^{143}a_{n}12^{n-y_{t}^{0,j-1}-6y_{t}^{1,j-1}-12y_{t}^{0,j}-72y_{t}^{1,j}}\right) \qquad \qquad y_{0}^{0,j}=\mathbf{d}_{6}(0,x_{0}^{j}) \label{thequaMou1b} \\
y_{t+1}^{1,j}&=&\mathbf{d}_{2}\left(0, \frac{1}{6}\sum_{n=0}^{143}a_{n}12^{n-y_{t}^{0,j-1}-6y_{t}^{1,j-1}-12y_{t}^{0,j}-72y_{t}^{1,j}}\right)  \qquad \quad y_{0}^{1,j}=\mathbf{d}_{2}(0,x_{0}^{j}/6) \label{thequaMou2b}\\
x_{t+1}^{j}&=&y_{t+1}^{0,j}+6y_{t+1}^{1,j}    \label{thequaMou3b}
\end{eqnarray}
If we put $n_{t}\equiv y_{t}^{0,j-1}+6y_{t}^{1,j-1}+12y_{t}^{0,j}+72y_{t}^{1,j}$ we note that if $a_{n_{t}} < 6$, $y_{t+1}^{1,j}=0$ and if $a_{n_{t}} \ge 6$ then $y_{t+1}^{1,j}=1$. By looking at the above table giving the 144 entries of the Moufang loop, we see that $y_{t+1}^{1,j}=0$ when $n_{t}$ takes values on the $6\times 6$ square blocks on the diagonal of the table. Conversely, we have that $y_{t+1}^{1,j}=1$ when $n_{t}$ falls in any of the nondiagonal square blocks. This in turn means that $y_{t+1}^{1,j}=1$ when either $y_{t}^{1,j-1}$ or $y_{t}^{1,j}$ are equal to one, but not both, \emph{regardless of the values of $y_{t}^{0,j-1}$ and $y_{t}^{0,j}$}. Thus the output $y_{t+1}^{1,j}$ is the XOR function (addition modulo 2 $+_{2}$ \cite{VGM2,VGM3}) of  $y_{t}^{1,j-1}$ or $y_{t}^{1,j}$ and we find that
\begin{equation}
y_{t+1}^{1,j}=\mathbf{d}_{2}\left(0, \frac{1}{6}\sum_{n=0}^{143}a_{n}12^{n-y_{t}^{0,j-1}-6y_{t}^{1,j-1}-12y_{t}^{0,j}-72y_{t}^{1,j}}\right)=y_{t}^{1,j}+_{2}y_{t}^{1,j-1}\equiv \mathbf{d}_{2}(0,y_{t}^{1,j}+y_{t+1}^{1,j-1}) \label{cuacuaqui}
\end{equation}
where the binary operator $+_{2}$ has table
\begin{center}
\begin{tabular}{c|cc}
 $\ +_{2} \ $ & $\ 0 \ $ & $\ 1 $ \\
\hline
$\ 0 \ $ & $\ 0 \ $ & $\ 1 \ $ \\
$\ 1 \ $ & $\ 1 \ $ & $\ 0 \ $ \\
\end{tabular}
\end{center}
So, the lifted layer corresponds to the simple CA with code $^{0}6^{1}_{2}$ \cite{VGM3} and rule vector $(a_{0},\ldots, a_{3})=(0,1,1,0)$, which is a \emph{layer CA}, independent of the ground layer. This rule is \emph{predictable}. For, let us define the shift operator $\mathbf{T}$ as
\begin{equation}
\mathbf{T}y_{t}^{1,j}=y_{t}^{1,j-1}
\end{equation}
it can be easily proved that the set of configurations of the CA forms a ring under addition modulo $2$ and the action of the shift operator. Therefore, since
\begin{equation}
y_{t+1}^{1,j}=y_{t}^{1,j}+_{2}y_{t}^{1,j-1}=(1+_{2}\mathbf{T})y_{t}^{1,j}
\end{equation}
starting from an arbitrary initial condition $y_{0}^{1,j}$ we obtain the solution as
\begin{eqnarray}
y_{t}^{1,j}&=&(1+_{2}\mathbf{T})^{t}y_{0}^{1,j}=\sum_{k=0}^{t}\binom{t}{k} \mathbf{T}^{t-k}y_{0}^{1,j} \mod 2 =\sum_{k=0}^{t}\binom{t}{k} y_{0}^{1,j+k-t} \mod 2 \\
&=&
\mathbf{d}_{2}\left(0,\sum_{k=0}^{t}\binom{t}{k} y_{0}^{1,j+k-t} \right) \label{predictable}
\end{eqnarray}
where the binomial theorem has been used. Thus, the solution at any later time $t$ is explicitly known from this equation for any arbitrary initial condition.

\begin{center}
\begin{figure*} 
\includegraphics[width=0.7 \textwidth]{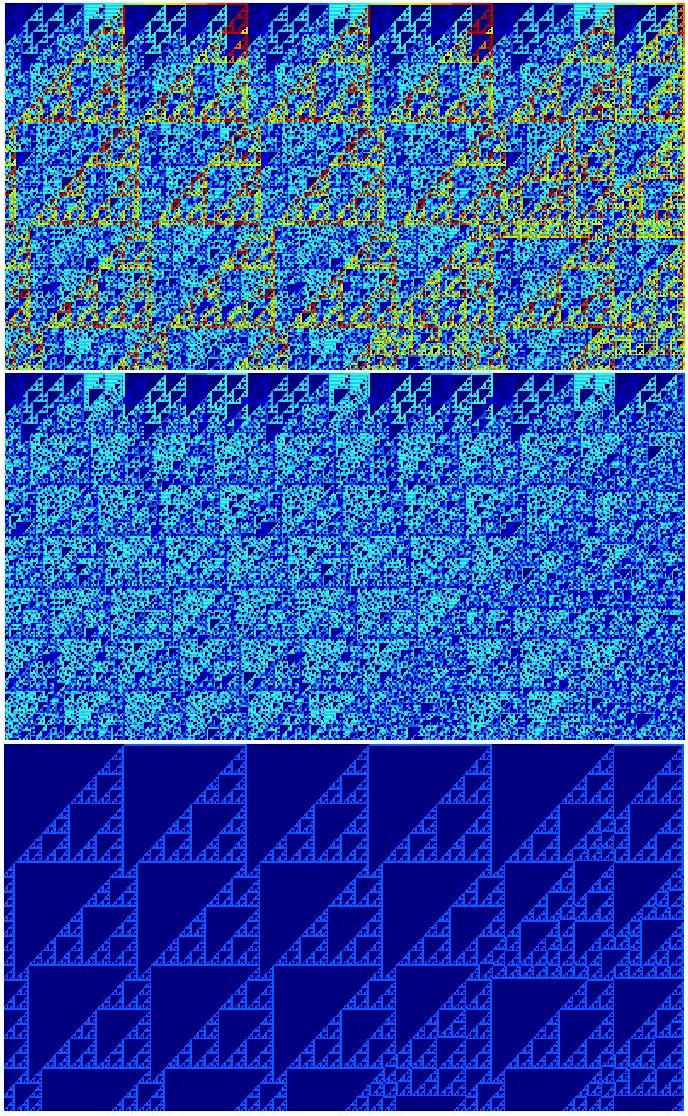}
\caption{\scriptsize{Spatiotemporal evolution of $x_{t}^{j}$ (top), $y_{t}^{0,j}=\mathbf{d}_{6}(0,x_{t}^{j})$ (center), and  $y_{t+1}^{1,j}=\mathbf{d}_{2}(1,x_{t}^{j}/6)$ (bottom) calculated from Eqs. (\ref{thequaMou1b}) to (\ref{thequaMou3b}) for the CA rule with $p=12$, $l=0$ and $r=1$ with rule vector given by Eq. (\ref{mouCA2}) and for an initial condition given by $x_{0}^{j}=\mathbf{d}_{12}(1,j)$. While the spatiotemporal evolution of the ground rule (center) is unpredictable, the one of the lifted layer corresponds to the simple CA $^{0}6^{1}_{2}$ which is predictable for any arbitrary initial condition. Time flows from top to bottom in each panel. Periodic boundary conditions are used, $j \in [1,400]$ and time $t \in [0,200]$.}}  \label{spatios}
\end{figure*}
\end{center}

The meaning of the layer described by Eq. (\ref{thequaMou2b}) which, as we have seen, reduces to Eq. (\ref{cuacuaqui}, with solution given by Eq. (\ref{predictable}) is now clear to explain. If one considers the sets of values $S_{0}\equiv \{0, 1, 2, 3, 4, 5\}$ and $S_{1}\equiv \{6, 7, 8, 9, 10, 11\}$ then Eq. (\ref{cuacuaqui}) specifies at each time to which set belongs $x_{t+1}^{j}$ from the information of those of $x_{t}^{j}$ and $x_{t}^{j-1}$. Equation (\ref{cuacuaqui})  establishes that, at time $t+1$ the value $x_{t+1}^{j}$ belongs to set $S_{1}$ if either $x_{t}^{j}$ or $x_{t}^{j-1}$ belong to $S_{1}$, but not both. A most remarkable fact is that \emph{we can exactly predict by means of Eq. (\ref{predictable}) at every time and for every initial condition $x_{0}^{j}$, to which set does $x_{t}^{j}$ belong}. 


We find, however that the ground layer \emph{is not} independent of the lifted layer and, hence, the Moufang rule is \emph{not} a graded rule. The behavior of Eq. (\ref{thequaMou1b}) is far more complex and no shortcut for the trajectory seems possible. Thus, the full evolution dictated by Eq. (\ref{thequaMou3}) is unpredictable as well. The conclusion of all this is that \emph{while we cannot predict the actual value of $x_{t}^{j}$ in this case, we can exactly predict at every time, from Eq. (\ref{predictable}), to which set $S_{0}$ or $S_{1}$ belongs the value of $x_{t}^{j}$}.  


 In Fig. \ref{spatios} we show the spatiotemporal evolution of $x_{t}^{j}$ (top), $y_{t}^{0,j}=\mathbf{d}_{6}(0,x_{t}^{j})$ (center) and $y_{t}^{1,j}=\mathbf{d}_{2}(1,x_{t}^{j}/6)$ (bottom) calculated from Eqs. (\ref{thequaMou1b}) to (\ref{thequaMou3b}) for an initial condition $x_{0}^{j}=\mathbf{d}_{12}(1,j)$.  We observe that, although for $x_{t}^{j}$ (top) and the ground layer $y_{t}^{0,j}$ (center) the spatiotemporal evolution is unpredictable, the one of the lifted layer $y_{t}^{1,j}$ (bottom) is exactly predictable and given by Eq. (\ref{predictable}) (we have numerically checked the correctness of this statement). This is so, of course, \emph{regardless of the initial condition}. The contribution of the lifted layer to the overall spatiotemporal dynamics is clearly discernible. Although the lifted layer is predictable and independent of the ground layer, the latter is dependent of the lifted layer and thus, the $p$-decomposable rule governing the overall dynamics is not a graded CA rule. Let us try to better understand the dynamics of the ground layer, which is obviously correlated to the lifted layer, as Fig. \ref{spatios} (center) and (bottom) show. Indeed, from Eq. (\ref{thequaMou1b}) there is an explicit dependence on the lifted layer through $y_{t}^{1,j}$ and $y_{t}^{1,j-1}$. We have now four possibilities: 1) $y_{t}^{1,j}=y_{t}^{1,j-1}=0$; 2) $y_{t}^{1,j}=0=1-y_{t}^{1,j-1}$; 3) $y_{t}^{1,j}=1=1-y_{t}^{1,j-1}$ and 4) $y_{t}^{1,j}=y_{t}^{1,j-1}=1$. In cases 1) and 2) the action of Eq. (\ref{thequaMou1b}) at time $t$ reduces to a CA $^{0}A^{1}_{6}$, in case 3) to a CA $^{0}B^{1}_{6}$, and in case 4) to another CA $^{0}C^{1}_{6}$. The rule tables of these CAs, which act on 6 symbols are given in the tables below
 \begin{center}
\begin{tabular}{c|cccccc}
 $\ ^{0}A^{1}_{6} \ $ & $\ 0 \ $ & $\ 1 $ & $\ 2 $ & $\ 3 $ & $\ 4 $ & $\ 5 $ \\
\hline
$\ 0 \ $ & $\ 0 \ $ & $\ 1 \ $ & $\ 2 \ $ & $\ 3\ $ & $ \ 4 \ $ &$\ 5 \ $  \\
$\ 1 \ $ & $\ 1 \ $ & $\ 0 \ $ & $\ 4 \ $ & $\ 5\ $ & $\ 2 \ $ &$\ 3 \ $ \\
$\ 2 \ $ & $\ 2 \ $ & $\ 5 \ $ & $\ 0 \ $ & $\ 4\ $&$\ 3 \ $&$\ 1 \ $ \\
$\ 3 \ $ & $\ 3 \ $ & $\ 4 \ $ & $\ 5 \ $ & $\ 0\ $&$\ 1 \ $&$\ 2 \ $ \\
$\ 4 \ $ & $\ 4 \ $ & $\ 3 \ $ & $\ 1 \ $ & $\ 2\ $&$\ 5 \ $&$\ 0 \ $ \\
$\ 5 \ $ & $\ 5 \ $ & $\ 2 \ $ & $\ 3 \ $ & $\ 1 \ $&$\ 0 \ $&$\ 4 \ $ \\
\end{tabular}
\qquad
\begin{tabular}{c|cccccc}
 $\ ^{0}B^{1}_{6} \ $ & $\ 0 \ $ & $\ 1 $ & $\ 2 $ & $\ 3 $ & $\ 4 $ & $\ 5 $ \\
\hline
$\ 0 \ $ & $\ 0 \ $ & $\ 1 \ $ & $\ 2 \ $ & $\ 3\ $ & $ \ 5 \ $ &$\ 4 \ $  \\
$\ 1 \ $ & $\ 1 \ $ & $\ 0 \ $ & $\ 4 \ $ & $\ 5\ $ & $\ 3 \ $ &$\ 2 \ $ \\
$\ 2 \ $ & $\ 2 \ $ & $\ 5 \ $ & $\ 0 \ $ & $\ 4\ $&$\ 1 \ $&$\ 3 \ $ \\
$\ 3 \ $ & $\ 3 \ $ & $\ 4 \ $ & $\ 5 \ $ & $\ 0\ $&$\ 2 \ $&$\ 1 \ $\\
$\ 4 \ $ & $\ 4 \ $ & $\ 3 \ $ & $\ 1 \ $ & $\ 2\ $&$\ 0 \ $&$\ 5 \ $ \\
$\ 5 \ $ & $\ 5 \ $ & $\ 2 \ $ & $\ 3 \ $ & $\ 1 \ $&$\ 4 \ $&$\ 0 \ $  \\
\end{tabular}
\qquad
\begin{tabular}{c|cccccc}
 $\ ^{0}C^{1}_{6} \ $ & $\ 0 \ $ & $\ 1 $ & $\ 2 $ & $\ 3 $ & $\ 4 $ & $\ 5 $ \\
\hline
$\ 0 \ $ & $\ 0 \ $ & $\ 1 \ $ & $\ 2 \ $ & $\ 3\ $ & $ \ 5 \ $ &$\ 4 \ $  \\
$\ 1 \ $ & $\ 1 \ $ & $\ 0 \ $ & $\ 5 \ $ & $\ 4\ $ & $\ 2 \ $ &$\ 3 \ $ \\
$\ 2 \ $ & $\ 2 \ $ & $\ 4 \ $ & $\ 0 \ $ & $\ 5\ $&$\ 3 \ $&$\ 1 \ $ \\
$\ 3 \ $ & $\ 3 \ $ & $\ 5 \ $ & $\ 4 \ $ & $\ 0\ $&$\ 1 \ $&$\ 2 \ $\\
$\ 4 \ $ & $\ 4 \ $ & $\ 2 \ $ & $\ 3 \ $ & $\ 1\ $&$\ 0 \ $&$\ 5 \ $ \\
$\ 5 \ $ & $\ 5 \ $ & $\ 3 \ $ & $\ 1 \ $ & $\ 2 \ $&$\ 4 \ $&$\ 0 \ $  \\
\end{tabular}
\end{center}
Thus, we reach the following conclusion: While the lifted layer is independent, the ground layer can be understood as being unfolded by the action of three different layer CAs of 6 symbols that act locally at $j$ and $t$. Which CA is chosen to act on that layer, depends on the values of  $y_{t}^{1,j}$ and $y_{t}^{1,j-1}$ of the lifted layer at $j$ and $t$. This is interesting because, in their course of action, complex systems frequently chose within alternatives as a function of other independent factors, and this process can be modeled by the CA above.

\section{Conclusions and perspectives}

In this article we have presented a method to decompose any nonlinear CA acting on an alphabet of $p$ symbols into a superposition of $N$ CAs, acting each on a number of symbols $p_{h}$, $h=0,\ldots, N-1$  that is a divisor of $p$. Since the prime factor decomposition of a number is unique (by the celebrated Euclid theorem) so is also that particular CA decomposition unique for each CA in rule space. This general result applies to \emph{all} CA in rule space and has lead us to discover a new class of dynamical systems that we have termed \emph{semipredictable dynamical systems}: In these, although the overall spatiotemporal dynamics is unpredictable, certain traits of the evolution can exactly be predicted. We have given explicit examples of semipredictable systems, from those involving elementary CA, to a rule of range $\rho=2$ acting on $p=12$ symbols with the structure of the smallest Moufang loop. We have seen that, in spite of the spatiotemporal dynamics being chaotic and unpredictable, the CA can be decomposed into two layers, one of them unpredictable and the other displaying a predictable spatiotemporal evolution/pattern for any initial condition. This is a remarkable fact since it shows that a chaotic scalar signal can carry with itself coherent information that can be extracted from a digital analysis of the signal.

The semipredictable dynamical systems presented here may find applications in the modeling of biological and complex systems. Most complex systems found in nature possess both predictable and unpredictable traits. For example, a certain disease on a complex organism can have a predictable evolution, with the dynamics of the complex organism that acts as host being unpredictable in many details that are relevant to the overall dynamics. By means of our model we have seen that, even when  predictable and unpredictable layers describing the dynamics are correlated, and hence not totally independent, they have sometimes robust features (closed substructures) that warrant their separate tractability and solvability. We give a summary of certain applications that seem naturally suited to the mathematical approach presented here.

\begin{itemize}
\item \emph{Physics:} Together with the $p\lambda n$ fractal decomposition that we have very recently introduced \cite{CHAOSOLFRAC}, graded CA might be applied to construct useful tensor products of wave functions in quantum mechanics. In fact, the connection of the digit function to Hilbert space is natural from our previous work  \cite{CHAOSOLFRAC} and from the principle of least radix economy \cite{QUANTUM}.  We have recently proposed the latter as a possible pathway to derive physical laws in a unified manner (encompassing both classical and quantum physics). The idea of semipredictable dynamical systems might mediate the quantum-classical correspondence in an analogous manner to the behavior found in this article for the ground and the lifted layers of the Moufang loop example. The key physical quantity to be addressed is the Lagrangian action $S$ \cite{QUANTUM} which is also  connected to entropy \cite{QUANTUM,Annals,Kalloniatis} and which yields a natural physical radix $\eta$ (which can be thought as related to $p$ in this paper) that establishes the character of the physical laws \cite{QUANTUM}.  Although we suggest that CA methods and fractality are of major importance in quantum mechanics, \emph{we do not assume any discrete or fractal spacetime at the Planck scale in \cite{QUANTUM}}, and \emph{we do not need such assumption in view also of the results reported in \cite{CHAOSOLFRAC}} where we show that fractals can be constructed out of continuous and differentiable functions that act as `mother functions'. Such latter complex-valued functions can be made Lorentz covariant and so shall then be any fractal objects derived from them through the method presented in \cite{CHAOSOLFRAC} as well. Other approaches based on CA \cite{Hooft,HooftNEW,Elze1,Elze2,Elze3} may also benefit from the mathematical methods introduced in this paper. Another application might be found in the analysis of complex fluids by means of lattice Boltzmann models \cite{Wolfram2,Chopard}: Lattice fluids involving many dynamical states can be decomposed in layers with the methods presented in this paper to get insight in their interactions and how the latter affect their spatiotemporal evolution.

\item \emph{Mathematics:} By considering cellular automata of neighborhood size $\rho=2$, the abstract algebraic properties of closed binary operators, also called laws of composition or magmas \cite{Bruck}, can be investigated by studying the spatiotemporal evolution of their associated CA rules. The Moufang loop presented here is an example. Finite groups and any other magmas \cite{Bruck,arxivCSo,CHAOSOLFRAC,arxiv4} can also be investigated in a similar manner. The idea of $p$-decomposability of the CA dynamics presented here is directly related to the Lagrange theorem of finite group theory. CA layers in graded CA rules are analogous to normal subgroups of finite groups. A graded CA rule can also be  understood as the dynamical counterpart of a direct sum of discrete operators. Of particular interest is also the connection established with multiplicative partitions \cite{Hughes}. Thus, our formalism provides an alternative pathway for the computational investigation of abstract algebraic structures. 


\item \emph{Complexity science:} Complementing our previous work \cite{VGM1,VGM2,VGM3,CHAOSOLFRAC,arxivCSo,VGM4} the mathematical method presented here might be further investigated in several directions within the realm of complexity science \cite{Wolfram,Holland,Crutchfield1,Crutchfield2,MooreBOOK,Barabasi1,Barabasi2,Strogatz1,Strogatz2}. In a previous work \cite{VGM3} we have introduced the concept of \emph{conditionally predictable dynamical systems} which, in brief, can be stated as those \emph{systems whose trajectory can be exactly predicted provided that they met no exception in tracing it}. In practice, a model for a conditional dynamical system is composed of a predictable part and a finite set of exceptions \cite{VGM3} that are introduced by means of suitable $\mathcal{B}$-functions \cite{VGM1, VGM3}. Thus, the trajectory is explicitly known provided that no exception has been met on it.  This, in turn, implies checking the whole trajectory for the occurrence of exceptions and, therefore, an associated degree of unpredictability and disbelief involved in the frailty of our prediction of the trajectory (if we renounce to carry out this check). By help of the $\mathcal{B}$-function, the semipredictable dynamical systems presented in this paper admit a straightforward generalization to \emph{conditionally semipredictable systems} when exceptions are added as a function of the dynamical states reached by the trajectory on each point in the discrete spacetime. The layers can also be coupled in complex manners by adding suitable $\mathcal{B}$-functions and realistic models for complex systems can be so designed for specific needs.
  

\item \emph{Music theory:} CA have previously been used in algorithmic music composition \cite{XenakisBOOK,MirandaBOOK} and we briefly sketch some ideas of how the methods presented in this article may contribute musical richness to this approach. In the theory of counterpoint \cite{TymoczkoBOOK} each of the different voices (instruments) evolves on a finite set of pitches forming an interesting melody that interweaves with the others. Each independent voice might be modeled by means of a suitable layer CA (probably of second or third order in time). Harmonic progressions, i.e. the sequences obtained from the evolution of simultaneous arrangements of the voices into chords, might thus be modeled by musically interesting graded CA rules. We conjecture that the latter might be constructed so as to evolve on Tymoczko orbifolds \cite{Tymoczko} thus warranting an efficient voice leading. A spatiotemporal evolution of such a CA would thus yield in parallel an ensemble  of musically consistent trajectories on these geometric structures. 


\item \emph{Biophysics and medical physics:} CA rules are often used as models for tumor growth and the spread of viruses and other diseases (see \cite{DeutschBOOK} and references therein). For example, in a very recent article \cite{Burkhead} a 2D stochastic CA model with alphabet size $p=4$ for the spread of the Ebola virus is considered. Although the CA there considered is stochastic, it is constructed out of three deterministic CAs working on four symbols and, hence, each being decomposable into two layers with $p_{0}=p_{1}=2$ symbols each.  Decomposable CA rules might be useful not only to get insight into realistic dynamical models, but also to help in the design of specific drugs in order to target specific layers of a living organism afflicted by a malignant agent, by leaving other layers unaffected, and to elucidate the way in which the malignant agent (which itself might be modeled by a $p$-decomposable CA rule) affects the organism as a whole.  

\item \emph{City landscapes and urban development:} Cities might be seen as complex multilayered structures \cite{Batty,BattyBOOK} emerging as a result of graded CA rules, some of them hierarchical. The mathematical methods presented in this paper, together with some others employed in the design of `exotic' fractal surfaces \cite{CHAOSOLFRAC,arxivCSo,Indekeu5,Barnsley}, might be applied to the understanding of these complex dynamical systems. 
\end{itemize}

We note that all results presented in this paper can be extended to two and three dimensions and that this aspect does not at all affect the conclusions on the decomposability of the CA rules: All these higher dimensional CAs can be described by means of suitable extensions of Eq. (\ref{themap}) by appropriately addressing the first argument of the digit function in Eq. (\ref{themap}) for the problem at hand.  We also note that layer CAs within a graded CA rule can also have different ranges (i.e. parameters $l$ and $r$ of the rules may be $h$ dependent). The only parameter that is relevant to warrant decomposability is the size of the alphabet $p$. After appropriate straightforward modifications, the CAs to be decomposed can also be of higher order in time.

We believe that the study of non-decomposable CA with $p$ a prime number (to which all Boolean CA with $p=2$ belong) is of fundamental interest, since they are the building blocks of all other CA dynamics in quite a similar way as prime numbers are building blocks for the non-negative integers. Further mathematical insights in non-decomposable CAs can be helpful to construct more subtle and increasingly realistic models of complex systems by means of graded CA rules, in particular, and $p$-decomposable rules, in general.

\bibliography{biblos}{}
\bibliographystyle{h-physrev3.bst}

\end{document}